\documentclass[printer]{aa}

\usepackage{graphicx}
\usepackage{txfonts}
\usepackage{natbib}
\usepackage[usenames,dvipsnames]{color}
\usepackage{mathrsfs}
\usepackage[fleqn]{mathtools}

\begin{document}

\title{A new nonlocal thermodynamical equilibrium radiative transfer method for cool stars}
\subtitle{Method \& numerical implementation}

\author
{
  J. Lambert    \inst{1,2}
  \and
  E. Josselin    \inst{2}
  \and
  N. Ryde         \inst{1}
  \and
  A. Faure              \inst{3}
}

\institute{
    {Lund Observatory, Box 43, SE-221 00 Lund, SWEDEN}
  \and
  {Laboratoire Univers et Particules de Montpellier (LUPM), UMR 5299,
    Universit\'e Montpellier 2 , CNRS \\
    Universit\'e Montpellier 2 - CC72 Place Eug\`ene Bataillon 34095 Montpellier Cedex 5 FRANCE}
  \and
  {Institut de Plan\`etologie et d'Astrophysique de Grenoble (IPAG), UMR 5109,
    Universit\'e Joseph Fourier, CNRS, OSUG, \\
    38041 Grenoble Cedex 9, FRANCE}
  \\
  \email{julien.lambert@astro.lu.se}            
}
\titlerunning{A new non-LTE Radiative Transfer Method for Cool Stars}

\date{Received <date>}

\abstract
{The solution of the nonlocal thermodynamical equilibrium (non-LTE) radiative transfer equation usually relies on stationary iterative methods, which may falsely converge in some cases. Furthermore, these methods are often unable to handle large-scale systems, such as molecular spectra emerging from, for example, cool stellar atmospheres. }
{Our objective is to develop a new method, which aims to circumvent these problems,  using nonstationary numerical techniques and taking advantage of parallel computers. 
}
{The technique we develop may be seen as a generalization of the coupled escape probability method. It solves  the statistical equilibrium equations in all layers of a discretized model simultaneously. The numerical scheme adopted is based on the generalized minimum residual method. }
{The code has already been applied to the special case of the water spectrum in a red supergiant stellar atmosphere. This demonstrates the 
fast convergence of this method, and opens the way to a wide variety of astrophysical problems. }
{}


\maketitle

\section{Introduction}

 Radiative transfer plays a central role in astrophysics. The solution of the radiative transfer equation (hereafter RTE) 
 is required to properly interpret any spectrum observed from an astrophysical object. Furthermore the radiation field often 
 represents an important ingredient in hydrodynamics, for instance, through the energy budget (radiative heating and/or cooling) 
 and the radiative pressure. However, as we discuss below, while the standard expression of the RTE is an 
 ordinary differential equation of the 1st order, it may be highly nonlinear and nonlocal, especially because of scattering 
 processes. The solution of the RTE thus requires specific procedures. Furthermore, with the development of infrared (IR) astronomy (e.g., from
  CRIRES@VLT in the near-IR to ALMA through HIFI/Herschel), one is now often confronted with 
 very large-scale problems resulting from molecular spectra, which most techniques developed so far cannot 
 handle. 

The techniques for solving the RTE may be divided into a few strategies. Among approximate methods, one that is widely used, relies on 
the escape probability technique \citep[e.g.,][]{Castor1970}, which does not take  the nonlocality of the RTE into account. 
Secondly,  Monte Carlo methods can be used because of the probabilistic nature of scattering processes. These methods can successfully be adapted to any geometry, but may be unadapted when large optical depths are met \citep[see, e.g., ][for a recent review]{Juvela2005}. We do not discuss these two classes of methods hereafter. 

To properly treat the nonlocality and the nonlinearity of the RTE, it is usually rewritten in the form of a diffusion integral 
equation, which is then solved iteratively\footnote{We already emphasize here that solving either the RTE explicitly or the statistical equilibrium equation is strictly identical, and both equations are of similar mathematical nature, as shown, for instance, in the Appendix of \cite{Hauschildt1995}}. This integral form is usually referred to as the $\Lambda$ operator.  
The form involves a kernel, which  determines the mathematical 
nature of the integral equation. For example, in the case of an assumed plane parallel geometry, the kernel involves the 
exponential integral function, and the RTE then becomes a weakly singular Fredholm equation of the second type 
(this singularity is an important aspect, which will be treated below).  It can then be formally solved thanks to appropriate 
algorithms, such as  the Atkinson algorithm \citep{Ahues2002}.Since the inversion of the $\Lambda$ operator is usually numerically prohibitive, one  needs an iteration scheme. It is well known 
that the iteration with the complete $\Lambda$ operator at best converges very slowly or even stabilizes far from the real solution. 
A major step in solving the RTE has been  the introduction of the so-called approximate lambda iteration (ALI), initially by 
\cite{Cannon1973}, then called the operator splitting method. The ALI is in fact an application of the Jacobi method to the RTE, as shown 
by \cite{Olson1986}, who greatly improved the mathematical understanding of the problem. 
In these methods, the choice of the approximate operator (denoted $\Lambda^\star$) is crucial. The most common choice for 
$\Lambda^\star$ is the diagonal part of $\Lambda$, which  ensures a rapid convergence if $\Lambda$ is diagonally 
dominant matrix (Gershgorin's Theorem). ALI has been applied to a large variety of problems, including 3D polarized transfer 
\citep{Stepan2013}. This latter work also emphasizes the potential progress that can be made by taking advantage of massively parallel computing. 

Meanwhile, new mathematical techniques have been applied to the RTE problem, especially the Gauss-Seidel method and the 
successive over relaxation (SOR) method, which significantly improve the convergence properties \citep[][]{Trujillo-Bueno:1995} that are determined by the spectral radius of their amplification matrix.
Besides these, complete linearization methods have also been developed, the well-known case being MULTI \citep{MULTI1985}. Despite apparent different formulations, the two approaches are essentially the same, as shown by \cite{Socas1997}.     

\begin{figure}
\centering
\includegraphics[width=\columnwidth]{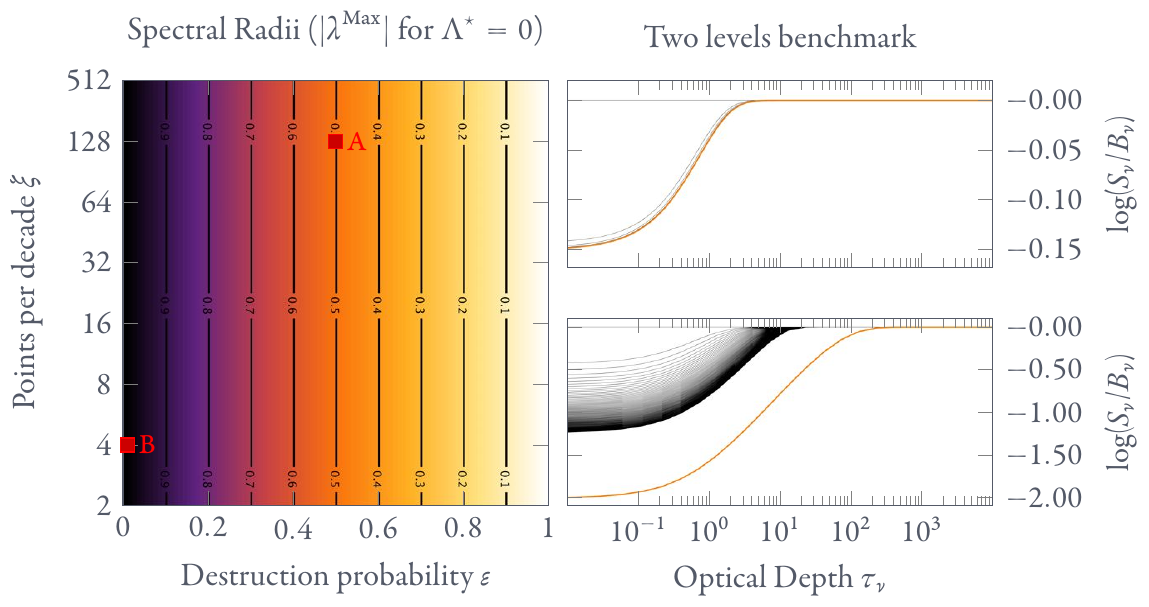}
\caption{Spectral radius of  amplification matrix for the $\Lambda$ iteration
method. Convergence is shown for two spectral radii, marked with letters 
A (upper panel) and B (lower panel) in the right-hand panels. The thin black lines correspond to successive iterations; the true solution is given by the orange line.}
  \label{fig:benchmark}
\end{figure}

\begin{figure}
\centering
\includegraphics[width=\columnwidth]{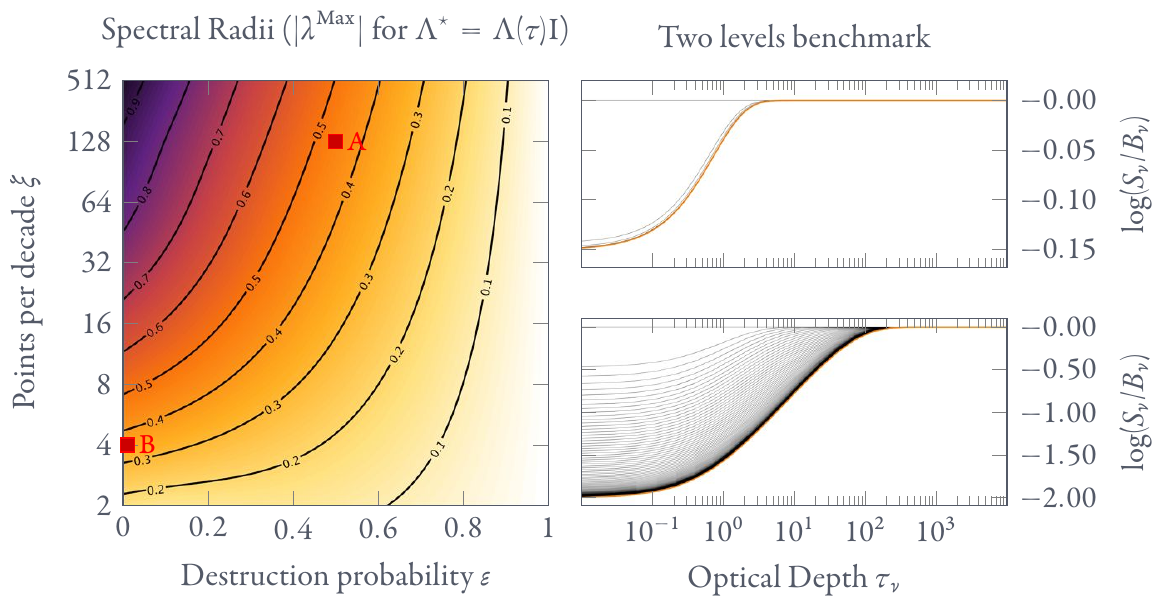}
\caption{Same as Figure \ref{fig:benchmark}, except  for ALI with $\Lambda^\star$ = diagonal
of $\Lambda$.}
\label{fig:benchmark2}
\end{figure}

Indeed, all these methods are based on linearization techniques. In particular, the dependence of the radiation field on 
the populations of the energy levels is approximated. Furthermore, all these methods are stationary (in the case of SOR, 
this is true if the relaxation parameter is constant over iterations), meaning that the spectral radius is constant. As the 
error vector $\mathbf{e}^m$ at the $m^{th}$ iteration is proportional to the power $m$ of the spectral radius of the 
amplification matrix $|\lambda_\mathrm{max}|^m$, the convergence may thus be very slow 
in some cases. For the Lambda iteration, the spectral radius is close to 1 when the photon destruction probability $\epsilon  \ll 1$ and/or the 
total optical depth is $\gg 1$, explaining why it falsely converges. This is illustrated in Figs. \ref{fig:benchmark} and 
\ref{fig:benchmark2}, where the classical Eddington problem (two-level atom with constant $\epsilon$) is solved 
with $\Lambda$ iteration (LI) and ALI, respectively. The false convergence of LI  and the very low convergence of ALI are 
illustrated in cases B. Indeed, their rate of convergence is very slow for strongly nonlocal thermodynamical equilibrium (non-LTE) problems (non-LTE parameter 
$\epsilon \ll 1$,  $\epsilon$ being the fraction of the source function given by the Planck function, or the photon 
destruction probability) 
and to the number of points per decade of optical depth. In the latter case (i.e., a fine sampling of the stellar atmosphere), 
this problem of slow convergence is well explained by \cite{FabianiBendicho1997}, who also showed that the 
multigrid method can nicely circumvent this problem. This fine sampling is however rarely required, especially as classical model atmospheres are static. An excellent discussion of the general problem of false convergence is given in 
Chapter 13 in \cite{Hubeny2014}. 

The iterative schemes are usually combined with a technique of acceleration of convergence \citep{Auer1991}. These fall into two 
main categories. The first  is based on the minimization of the residual, which is the case of the Ng acceleration \citep{Ng1974}. The 
second class relies on minimization with respect to a set of conjugate vectors. This was first applied to the RTE by (\citealt{Klein1989} ; see also \citealt{Dickel1994}).

The second class of techniques (the minimization with respect to a set of conjugate vectors) leads naturally to more general, nonlinear, nonstationary methods, such as the conjugate gradient and the 
generalized minimum residual method (GMRES). To our knowledge, only the conjugate gradient has been applied to the RTE, 
by \cite{Paletou2009}. 

The aim of our work  is to develop a new nonstationary method for solving the RTE, which fully takes 
 its nonlinearity into account. This new method thus opens the possibility to explore the mathematical consequences of this intrinsic property 
of the RTE. Furthermore, circumventing the possible slow or false convergence of stationary methods should allow an 
appropriate treatment of some specific cases, such as, e.g., shocked regions, which require a fine sampling and present a large dynamics of optical depth values. Finally, as the coding strategy of this method has been conceived as a parallel code 
from the beginning, it is applicable to large-scale systems such as molecular spectra, and this was indeed our initial motivation
(a posteriori parallelization is often less efficient, or even impossible).

The derivation of the equations, including the Jacobian matrix coefficients, used in the nonlinear method is 
presented in section 2. The implementation of this method and its parallelization are described in section 3. The application of the code 
to a specific problem facilitates the determination of its convergence properties, as shown in section 4. 


\section{Global strategy}

Contrary to the two-level atom problem with a constant destruction probability, which can be directly  solved 
by  direct inversion (or through the analytic Eddington solution). Multi-level atom or molecule problems require an iterative determination 
for two reasons.  Firstly, the multi-level problem is nonlinear. Secondly, the radiative transfer equation 
is nonlocal, and one has to propagate the modification of the field during iterations. As we discussed in the previous section, however, this stationary iteration may be problematic for large-scale systems.
The basic concept of the strategy we adopt is to suppress the iterations due to the nonlocality of the RTE by considering the full system 
(spatially) and computing the radiation field exactly (assuming given populations) with an analytic formulation of the 
mean intensity field, which explicitly depends on populations.

\subsection{Formulation of the problem }

For clarity, we hereafter  consider the special case of plane parallel geometry. In this context the 1D RTE is
\begin{equation}
  \label{eq:Radiativ transfer equationCH2}
  \mu \frac{dI_\nu}{ds}=\eta_\nu^\mathrm{Cont}+\eta_\nu^\mathrm{Line}-(\chi_\nu^\mathrm{Cont}+ \chi_\nu^\mathrm{Line})I_\nu
,\end{equation}
where $I_\nu$ is the specific intensity, $\eta$ and $\chi$ are the emission and extinction coefficients, 
 and $s$ is the geometrical path. 
We explicitly separate the continuum and the line processes (all demonstrations and symbols can be found in Appendix \ref{appA} and \ref{appC}, respectively). Assuming complete redistribution, we introduce classical expressions for line processes,
\begin{subequations}%
  \begin{align}
    \label{eq:def1a}
    \chi_\nu^\mathrm{Line} & = \phi^{ul}_\nu  \chi_{ul} = \phi^{ul}_\nu  \frac{h{\nu_{ul}}}{4\pi}  (B_{lu}n_{l}-B_{ul}n_{u})\\
    \label{eq:def1b}
    \eta_\nu^\mathrm{Line} & = \phi^{ul}_\nu \eta_{ul} =  \phi^{ul}_\nu \frac{h{\nu_{ul}}}{4\pi}A_{ul}n_{u,}
  \end{align}%
\end{subequations}
where $u$ and $l$ refer to the upper and lower levels, respectively. 
Hereafter, if the distinction between 
upper and lower levels is not required, the indices $i$ and $j$ are used. The normalized line profile is  
$\phi^{ul}_\nu$   and $A$ and $B$ are  the Einstein coefficients.

Following the strategy of \cite{Gonzalez-Garcia:2008}, we  then introduce a unique optical depth scale for
all frequencies, e.g,. the commonly used $\tau_\mathrm{V}$ scale 
(optical depth at 500 nm).
Eq. \eqref{eq:Radiativ transfer equationCH2} then becomes
\begin{equation}
  \label{eq:Radiativ transfer equationvar}
  \mu\frac{dI_\nu}{d\tau_\mathrm{V}}=\left[ \xi_\nu + E_{ij}\zeta\phi^{ij}_\nu(x_j-x_i)\right]I_\nu - D_{ij}x_i\zeta\phi^{ij}_\nu-\xi_\nu S_\nu
,\end{equation}
where 
\begin{subequations}
  \label{eq:def_xi}
  \begin{align}
\xi_{\nu} & =\frac{\tilde{\chi}_{\nu_{ij}}^\mathrm{Cont}}{\tilde{\chi}_\mathrm{V}} \\
\zeta & =\frac{1}{\tilde{\chi}_\mathrm{V}}\frac{n^{k}}{n^{H}} \\
E_{ij} & =\frac{h\nu_{ij}}{4\pi}B_{ij}g_i \\
D_{ij} & =\frac{h\nu_{ij}}{4\pi}A_{ij}g_i \\
x_i & =\frac{n_i}{n^k}\frac{1}{g_i}. 
  \end{align}%
\end{subequations}
The notation $\tilde{.}$ means that the quantity is  expressed per hydrogen atom.
Details can be found in Appendix \ref{appA}.
As implicitly assumed in Eq. \ref{eq:Radiativ transfer equationvar}, we do not consider line overlap. This assumption is justified in the case of infrared molecular lines. Otherwise, a summation over transitions (at least adjacent ones) 
would be required. 

This formulation depends explicitly on the populations of each energy level of the species considered. These populations are solutions of the statistical equilibrium equation system \footnote{At steady-state. Chemical formation and destruction terms are not included.},
which is for each level $n_i$, the detailed balance of incoming and outgoing (de-)excitation processes, defined by radiative (Einstein coefficients), and collisional rates 
($n_{col}$ being the local density of the colliding partner). Thus the statistical equilibrium equations take the form
\begin{equation}%
  \label{eq:ES}
  \begin{split}
    \frac{dn_i}{dt} = n_i & \left[\sum_{j<i}{A_{ij}+
        \sum_{j\ne i}{\left(B_{ij}{\bar{\mathcal{J}}}_{ij}+
            n_{col} C_{ij} \right)}}\right]\\
    - & \left[\sum_{j>i}{n_j A_{ji}+
        \sum_{j\ne i}{n_j \left(B_{ji}{\bar{\mathcal{J}}}_{ji}
            +n_{col}C_{ji} \right)}}\right]=0
  \end{split}
,\end{equation}
where the mean radiation field ${\bar{\mathcal{J}}}_{ij}({\tau }_V)={\bar{\mathcal{J}}}_{ji}({\tau }_V)$ depends on the solution of the RTE,
\begin{equation}%
  \label{eq:jbar}%
  {\bar{\mathcal{J}}}_{ij}(\tau_\mathrm{V}) = \frac{1}{2} \int_{0}^{\infty}{ \phi^{ij}_\nu (\tau_\mathrm{V})
    \int_{-1}^{+1} {I_\nu(\tau_\mathrm{V},\mu) d\mu} d\nu}
.\end{equation}%
Because this system of equations is degenerate, one of them is replaced by the normalization equation $\sum_i^N{n_i}=n^k$, where $n^k$ is the total density of the species $k$.

To express each equation of the statistical equilibrium system with the same unknowns as in Eq. \ref{eq:Radiativ transfer equationvar}, we rewrite the statistical equilibrium with the  variables $x_i$ defined in Eq. \ref{eq:def_xi}. Moreover, for numerical reasons, we make a variable change for the mean intensity,
\begin{equation}
\label{eq:defZ}
{\bar{\mathcal{J}}}_{ij}(\tau_\mathrm{V}) = \frac{2h\nu_{ij}^3}{c^2}{\bar{\mathcal{Z}}}_{ij}(\tau_\mathrm{V}) 
.\end{equation}
This leads to a matrix formulation of the statistical equilibrium, with an upper triangular part corresponding to incoming processes, and 
a lower triangular part corresponding to outgoing processes. Finally, thanks to the Einstein relations 
($B_{ul}=(c^2/(2h\nu_{ul}^3))A_{ul}$ 
and $B_{lu}= (g_u/g_l) B_{ul}$), the system of equations \eqref{eq:ES} becomes
\begin{equation}%
  \label{eq:ES2}
  \begin{split}
    x_i & \left[\sum_{j<i}{g_i {A}_{ij}\left(1+{\bar{\mathcal{Z}}}_{ij}\right)} +
    \sum_{j>i}{g_j {A}_{ji}{\bar{\mathcal{Z}}}_{ij}} +
        \sum_{j \ne i}{n_{col} g_i {C}_{ij}}\right]\\
    -& \left[\sum_{j<i}{x_j g_j {A}_{ij}{\bar{\mathcal{Z}}}_{ij}} +
    \sum_{j > i}{x_j g_j {A}_{ji}\left(1+{\bar{\mathcal{Z}}}_{ji}\right)} +
        \sum_{j \ne i}{n_{col} g_i {C}_{ij}}
    \right]=0
  \end{split}
.\end{equation}
Because radiative transfer is a nonlocal problem, an integral formulation of the mean radiation field with an 
explicit dependence on populations is required.

\subsection{The mean radiation field}
To derive the mean radiation field, the first step consists of deriving the formal
solution of the RTE. A classical way to obtain an integral form of the formal solution  of a differential equation is to use 
the Green's functions $\mathscr{G}_{\mu \nu}$. This method consists of replacing the source term by a Dirac function. Taking the boundary conditions  into account, 
the equation is then solved using its Laplace transform. The complete solution is then obtained after integration of this 
Green's function, that is, performing a summation over each point source, according to the superposition principle (which is valid if the 
differential operator is linear). Thus, the integral form of $I_{\nu}(\tau_\mathrm{V},\mu)$ can be obtained by the determination of the
Green's function of Eq. \eqref{eq:Radiativ transfer equationvar}. This Green's function is easily computable for spherical and other geometries using some symbolic tools such as Mathematica, making it possible to generalize the method for 2D or 3D problems. Knowing this Green's function, we can directly obtain the formal integral form of the specific intensity, which is
\begin{equation}%
  \label{eq:Inu}%
  I_{\nu}(\tau_\mathrm{V},\mu)=%
  \int^{\tau^\mathrm{Max}_\mathrm{V}}_{\tau^\mathrm{Min}_\mathrm{V}}{\left[
      D_{ij}x_i(t)\zeta(t)\phi^{ij}_\nu(t)
      +\xi_\nu(t) S_\nu(t)\right]\mathscr{G}_{\mu \nu}(\tau_\mathrm{V};t)dt}
.\end{equation}
We assume the line profile does not  depend on $\mu$. This expression is thus not applicable in the case 
of anisotropic scattering or  in the presence of radiation fields, but is valid in the case of classical static model atmospheres.
The formal expression of the mean intensity, ${\bar{\mathcal{J}}}_{ij}({\tau }_\mathrm{V})$, requires an integration over solid angles and frequencies according to Eq \eqref{eq:jbar}. We assume that $D_{ij}x_i\zeta$ and $\xi_{\nu}S_{\nu}$ are angular independent (physically this corresponds to the assumption of isotropic scattering) and frequency independent over the line width ($\xi_{\nu}S_{\nu} \rightarrow  \xi_{ij}S_{\nu_{ij}}$). This is a very weak assumption because the continuum is nearly constant on the scale of a line width. Then, we can extract the source terms from the integral over solid angles and frequencies. Moreover, we separate line and continuum contributions by splitting the integral, which leads to
\begin{equation}%
  \begin{split}%
    {\bar{\mathcal{J}}}_{ij}({\tau }_\mathrm{V})
    &=\int^{{\tau}^\mathrm{Max}_\mathrm{V}}_{{\tau
      }^\mathrm{Min}_\mathrm{V}}{\xi_{{ij}}(t) S_{\nu_{ij}}(t)}\\
    \times & \left[\frac{1}{2}\int_{0}^{\infty}{ \phi^{ij}_\nu
        (\tau_\mathrm{V}) \left[\int_{-1}^{+1}{ \mathscr{G}_{\mu
              \nu}(\tau_\mathrm{V};t)d\mu}\right]d\nu}\right]dt\\
    &+D_{ij}\int^{{\tau}^\mathrm{Max}_\mathrm{V}}_{{\tau}^\mathrm{Min}_\mathrm{V}}{x_i(t)\zeta(t)}\\
    \times & \left[\frac{1}{2}\int_{0}^{\infty}{ \phi^{ij}_\nu
        (\tau_\mathrm{V}) \phi^{ij}_{\nu}(t) \left[
          \int_{-1}^{+1}{\mathscr{G}_{\mu
              \nu}(\tau_\mathrm{V};t)d\mu}\right]d\nu}\right]dt
  \end{split}
.\end{equation}
Then, after determination of the Green's function in a given geometry, one obtains an integral form for the different
contributors to the radiation field. We assume that the continuum contribution is in LTE $(S_{\nu_{ij}}=B_{\nu_{ij}}(T))$. 
Moreover, we consider, as boundary conditions, 
an incoming field $I_-^\mathrm{Ext}$ from the inner atmosphere. 
Then, 
\begin{equation}
    \bar{\mathcal{J}}_{ij}(\tau_\mathrm{V})=
    \bar{\mathcal{J}}_{ij}^\mathrm{Ext}(\tau_\mathrm{V})
    +\bar{\mathcal{J}}_{ij}^\mathrm{Cont}(\tau_\mathrm{V})
    +\bar{\mathcal{J}}_{ij}^\mathrm{Line}(\tau_\mathrm{V})
\end{equation}%
where, 
\begin{subequations}%
    \label{eq:Jtot}
        \begin{align}%
      \bar{\mathcal{J}}^\mathrm{Ext}_{ij}[x_i,x_j] (\tau_\mathrm{V})%
      =&I_-^\mathrm{Ext}\beta_-[x_i,x_j](\tau_\mathrm{V})
      \label{eq:Jtota} \\%
      \bar{\mathcal{J}}^\mathrm{Cont}_{ij}[x_i,x_j] (\tau_\mathrm{V})%
      =&\int^{{\tau}^\mathrm{Max}_\mathrm{V}}_{{\tau}^\mathrm{Min}_\mathrm{V}}{\xi_{{ij}}(t)B_{\nu_{ij}}(T(t))%
        L_1 [x_i,x_j](t,\tau_\mathrm{V})dt}%
      \label{eq:Jtotb} \\%
      {\bar{\mathcal{J}}}^\mathrm{Line}_{ij}[x_i,x_j] (\tau_\mathrm{V})%
      =&D_{ij}\int^{{\tau}^\mathrm{Max}_\mathrm{V}}_{{\tau}^\mathrm{Min}_\mathrm{V}}{x_i(t)\zeta(t)%
        K_1 [x_i,x_j](t,\tau_\mathrm{V})dt,}%
      \label{eq:Jtotc}%
    \end{align}%
  \end{subequations}%
where $\beta_-[x_i,x_j](\tau_\mathrm{V})$ is the escape probability from position $\tau_\mathrm{V}$ in  the atmosphere,
 \begin{equation}
    \label{eq:beta}%
        \beta_-[x_i,x_j](\tau_\mathrm{V}) =  \frac{1}{2}\int_{0}^{+\infty}{\phi^{ij}_\nu(\tau_\mathrm{V})}\\
        \times { E_2 \left(\tau^\mathrm{Tot}[x_i,x_j]
          (\tau_\mathrm{V})\right) d\nu}
 .\end{equation}

The kernels $L_1 [x_i,x_j](t,\tau_\mathrm{V})$ and $K_1 [x_i,x_j](t,\tau_\mathrm{V})$ are, respectively: 
\begin{subequations}%
    \label{eq:kernel}%
    \begin{align}%
      \begin{split}%
        L_1&[x_i,x_j](t,\tau_\mathrm{V}) =%
        \frac{1}{2}\int_{0}^{+\infty}{\phi^{ij}_\nu(\tau_\mathrm{V})}\\%
        & {\times  E_1\left(-\varepsilon(t,\tau_\mathrm{V})(\tau^\mathrm{Tot}[x_i,x_j](t)%
            -\tau^\mathrm{Tot}[x_i,x_j](\tau_\mathrm{V}))\right) d\nu}%
      \end{split}\\%
      \begin{split}%
        K_1&[x_i,x_j](t,\tau_\mathrm{V}) =%
        \frac{1}{2}\int_{0}^{+\infty}{\phi^{ij}_\nu(\tau_\mathrm{V})\phi^{ij}_\nu(t)}\\%
        & {\times E_1\left(-\varepsilon(t,\tau_\mathrm{V})(\tau^\mathrm{Tot}[x_i,x_j](t)%
            -\tau^\mathrm{Tot}[x_i,x_j](\tau_\mathrm{V}))\right) d\nu,}%
      \end{split}%
    \end{align}%
  \end{subequations}%
where the total optical depth is given by
\begin{equation}
\begin{split}
\tau_\nu^\mathrm{Tot}(\tau_\text{V}) & =\int_{0}^{\tau_\text{V}}{ \left( \chi_\nu^\text{Cont}+ \chi_\nu^\text{Line} \right) dt} \\
& =\int_{0}^{\tau_\mathrm{V}}{\left(\xi_\nu(t) + E_{ij}\zeta(t)\phi^{ij}_\nu(t)[x_j(t)-x_i(t)]\right) dt}
\end{split}
.\end{equation}
We do not include an external radiation field. In the plane parallel geometry considered here, it would require  us
to take  additional effects such as limb darkening in realistic cases into account (e.g., illumination by a planet or a companion). 
We have introduced the function $\varepsilon(t,\tau_\mathrm{V}) = 1 \text{ if } t < \tau_\mathrm{V} \text{ and } -1 \text{ if } t > \tau_\mathrm{V} $  to split the integration over $\mu$ according to its sign. 
The physical meaning of the kernels $L_1(t,\tau_\mathrm{V})$ and $K_1(t,\tau_\mathrm{V})$ can be understood as  the probability density  
for photons, emitted in a layer $t$, to contribute to the line excitation in the layer $\tau_\mathrm{V}$. \\
We now have an analytic expression of the mean intensity with an explicit dependence on populations. We now have to include this expression in the equations of statistical equilibrium to close the system and form a system of coupled integral equations (a Fredholm-like equation system).

\subsection{The multi-zone statistical equilibrium equation system}
The inversion of the system of statistical equilibrium equations allows us to obtain the populations also including the formal solution of RTE. 
But because it is nonlocal, the knowledge of $\bar{\mathcal{J}}$ requires the knowledge 
of the populations everywhere in the atmosphere because of the integral over
the optical depth
 $t$ 
in Eq. \eqref{eq:Jtot}.

For this reason, one needs to solve the statistical equilibrium everywhere in the atmosphere simultaneously. 
The first step consists of rewriting Eq. \eqref{eq:ES} in a matrix form. A vector
$\mathbf{x}({\tau }_V)$ 
containing all the populations of the considered species in a given layer, normalized to the density in that layer, is 
built. The size of this vector is thus equal to $N$, the number of energy levels of the considered species. 
We introduce the rate matrix
\begin{equation}%
    \label{eq:matriceR}%
    \begin{split}
    \mathbf{R}[ \mathbf{x}]({\tau }_\mathrm{V}) &=%
    \mathbf{A}^T \circ \left( \mathbf{L}_1 +\bar{\mathbf{Z}}[\mathbf{x}]({\tau }_\mathrm{V})\right)^T \\
    &+\mathrm{diag(\mathbf{g})} \cdot \left[\mathbf{A}\circ \bar{\mathbf{Z}}[\mathbf{x}]({\tau }_\mathrm{V})\right] \cdot \mathrm{diag(\mathbf{g})}^{-1}+\mathbf{C}^T({\tau }_\mathrm{V})%
    \end{split}
.\end{equation}
Where $\mathbf{L}_1$ is a lower triangular matrix of ones, 
$\mathbf{A}$ is a vector containing the 
spontaneous de-excitation coefficients ($A_{ul}$), $\mathbf{Z}$ is a matrix containing the radiation field  
(see Eq. \ref{eq:defZ}), $\mathbf{C}$ a matrix containing the collisional rates 
($n_{col} C_{ij}$), and $\mathbf{g}$ is a vector containing the statistical weights ($g_i$).\\

Now the statistical equilibrium  system can be written as,
\begin{equation}%
    \label{eq:ES4}%
    \mathbf{M}^{T}[ \mathbf{x}]({\tau }_\mathrm{V}) \mathbf{x}({\tau }_\mathrm{V}) = \mathbf{0}%
,\end{equation}%
with
\begin{equation}%
    \label{eq:matriceM}%
    \mathbf{M}[\mathbf{x}](\tau_\mathrm{V})=\left[\mathrm{diag(\mathbf{u} \cdot \mathbf{R}[ \mathbf{x}]({\tau }_\mathrm{V}))} - \mathbf{R}[ \mathbf{x}]({\tau }_\mathrm{V})\right] \cdot \mathrm{diag(\mathbf{g})}
  ,\end{equation}%
where $\mathbf{u}$ is a ones vector.
After discretization of the medium into $P$ layers indexed by $l$, the vectorial function 
$\mathbf{x}(\tau_\mathrm{V})$ becomes $\mathbf{x}^l$ and equation \eqref{eq:matriceR} becomes
\begin{equation}%
    \begin{split}
    &\mathbf{R}^l(\mathbf{x}^1,\mathbf{x}^2,...,\mathbf{x}^P) =%
    \mathbf{A}^T \circ \left( \mathbf{L}_1 +\bar{\mathbf{Z}}^l(\mathbf{x}^1,\mathbf{x}^2,...,\mathbf{x}^P)\right)^T \\
    &+\mathrm{diag(\mathbf{g})} \cdot \left[\mathbf{A}\circ \bar{\mathbf{Z}}^l(\mathbf{x}^1,\mathbf{x}^2,...,\mathbf{x}^P)\right] \cdot \mathrm{diag(\mathbf{g})}^{-1}+\left[\mathbf{C}^{l}\right]^T%
    \end{split}
\end{equation}%
as the radiation field, and thus the matrix $\mathbf{Z}$ depends implicitly on the populations in all layers
, and 
the statistical equilibrium system can be rewritten,%
\begin{equation}%
    \label{eq:MatEqStat}%
    \left[\mathbf{M}^l(\mathbf{x}^1,\mathbf{x}^2,...,\mathbf{x}^P)\right]^{T}\mathbf{x}^l = \mathbf{0}%
  .\end{equation}%
To take care of the coupling between the layers and close the system  to solve Eq. \eqref{eq:MatEqStat} for all $l$ in a    %
unique step, we build the multi-zone statistical equilibrium system:
\begin{equation}%
    \label{eq:MatEqStat2}%
    \left( \begin{array}{ccc}%
        {{\mathbf M}}^1\left({\mathbf X}\right) & 0 & 0\\%
        0 & \ddots  & 0\\%
        0 & 0 & {{\mathbf M}}^p\left({\mathbf X}\right) \end{array}%
    \right)^{T}%
    \left( \begin{array}{c}%
        \mathbf{x}^1\\%
        \vdots \\%
        \mathbf{x}^P%
      \end{array} \right)=%
    \left( \begin{array}{c}%
        \mathbf{0} \\%
        \vdots \\%
        \mathbf{0}%
      \end{array} \right)%
  .\end{equation}%
We introduce the vector $\mathbf{X}=(\mathbf{x}^1,\mathbf{x}^2,...,\mathbf{x}^P)$ and the block-diagonal     %
matrix ${\mathbf \Gamma}$ to rewrite Eq. \eqref{eq:MatEqStat2} as a multi-D nonlinear function,
\begin{equation}%
    \label{eq:fonction}%
    \left[\boldsymbol{\Gamma}(\mathbf{X})\right]^{T}\mathbf{X}=\mathbf{F}(\mathbf{X})=\mathbf{0}%
  .\end{equation}%
With an initial $\textbf{X}_0$ within the convergence radius, we can solve this nonlinear equation with a  classical scheme such as the  Newton method.
An important question, however, is to determine whether the above system has a solution and whether the solution we expect to get is physical (unique and positive).

\subsection{Existence, uniqueness and positivity of the solution} 

From a physical point of view, one naturally expects a unique and positive solution of the RTE. Reflecting on
this problem is in fact a way to verify whether it is mathematically well formulated. Indeed, as stated
above, the coupling between the populations and the radiation field is nonlinear. 
Without linearization, the solution can be regarded as the equilibrium state of Markov
processes. Furthermore, because of radiative selection rules and collisional propensity rules, many factors
are very small, and may numerically tend toward zero. The problem is thus potentially ill-conditioned. 

The formulation adopted here is a multi-zone statistical equilibrium. The problem of uniqueness and
positivity of the solution of the mono-zone statistical equilibrium has been addressed by
\cite{Damgaard:1992} and \cite{Rybicki:1997}. As this system of equations is degenerate, one equation is
usually replaced by a normalization condition, namely  the conservation equation, to overcome the   
singularity of the associated matrix. Then, \cite{Damgaard:1992} show that the solution is positive {\it \textup{if
all the rates are strictly non-zero}}. Yet certain transition probabilities vanish, physically or numerically.
As long as all the levels are connected, just one nonzero cofactor guarantees the regularity of the
matrix. More generally, \cite{Rybicki:1997} demonstrated that even for the system without a normalization
condition, positive states can link (and be linked) only to positive states. Then, if the statistical
equilibrium equations can be transformed into a set of uncoupled irreducible subproblems with a strictly
positive normalization condition, the solution is positive and unique. In other words, the uniqueness and
positivity properties of the general \textup{\textup{{\it linear} }}statistical equilibrium equations relate to the underlying
connectivity properties of the states.

In our multi-zone formulation, the coupling between statistical equilibria in every atmospheric layer is
strongly nonlinear. That is, the system cannot be split into uncoupled sets of equations, i.e., a set of
uncoupled irreducible subproblems. This may impact the number of required normalization conditions, while
only one per atmospheric layer (the conservation equation) is available. The uniqueness of the solution may
then be demonstrated through the search for a nonlinear fixed point as the solution of the equilibrium of a
Markov chain (P. Azerad, priv. comm.). A forthcoming paper will be devoted to this. At least, as the properties
of connectivity  still apply, the positivity of the solution is guaranteed. 

\section{Implementation}
\subsection{Nonlinear solving}
\label{solving}

In the method developed here, the computation of the statistical equilibrium is based on its exact formulation 
and explicitly includes radiative transfer effects through the mean radiation field $\bar{\mathcal{J}}$ in equation 
\eqref{eq:fonction}. 
Finding the root of this function naturally leads to the exact solution to the statistical equilibrium, and avoids any false 
convergence that may be obtained with stationary iterative methods. 

The computation of the norm of a function,
\begin{equation}
 \mathbf{F}:\left({\mathbb{R}^{+}}\right)^{N \times P} \rightarrow \mathbb{R}^{N \times P}
,\end{equation}
$N$ and $P$ being the number of considered energy levels and the number of layers in the model atmosphere, respectively, is equivalent to an optimization problem, that is, the search for the minimum of the $L_2$ norm of the function $f$,
\begin{equation}
\min_{\mathbf{X} \in  \mathbb{R}^{N\times P}} f(\mathbf{X})=\frac{1}{2} \textbf{F}(\mathbf{X})\textbf{F}^{T}(\mathbf{X})=\frac{1}{2}{\|\textbf{F}(\mathbf{X})\|}^2_{L_2}
.\end{equation}

The classical way to minimize this function is to move along a vector $\delta \textbf{X}$ in a descending direction, 
using the Newton's method. Indeed, for $\delta \textbf{X} = \textbf{X}_k - \textbf{X}_{k-1} = -\textbf{J}^{-1} \cdot \textbf{F}$, 
the gradient along the vector 
$\delta \textbf{X}$
is
\begin{equation}
\label{eq:opti}
\begin{split}
\nabla f(\mathbf{X}_k)^{T} \delta \textbf{X} &= \left[\textbf{F}(\mathbf{X}_k)^{T} \textbf{J}(\mathbf{X}_k)\right]\left[-\textbf{J}(\mathbf{X}_k)^{-1}\textbf{F}(\mathbf{X}_k)\right] \\
& = -\textbf{F}(\mathbf{X}_k)^{T} \textbf{F}(\mathbf{X}_k) = -{\|\textbf{F}(\mathbf{X})\|}^2_{L_2} < 0
\end{split}
.\end{equation}

Newton's method is very efficient in approaching the solution as it has a quadratic convergence.
For each successive Newton step, we have to solve the linear problem $\textbf{J} \delta \textbf{X}_k = -\textbf{F}$, where $k$ is the index of the Newton iteration. 
However, with classical algebric methods, this operation leads to an $\mathcal{O}(N^3)$ algorithm, which is prohibitive  
in terms of computational time for large-scale systems. A robust and efficient way to accelerate this operation is to approximate the solution using a faster algorithm. We adopt GMRES, which is a subclass of Krylov subspace methods. 

A Krylov subspace method 
begins with an initial guess $\delta \textbf{X}_k^0$. At the $n^\text{th}$ sub-iteration, $\delta \textbf{X}_k^n$ is determined through a 
correction in the $n^\text{th}$ Krylov subspace,

\begin{equation}
\mathcal{K}_k^n = \text{span}(\textbf{R}, \textbf{J}\textbf{R}, . . ., \textbf{J}^{n-1}\textbf{R}) ; \textbf{R} = -\textbf{F} - \textbf{J} \delta \textbf{X}_k^n
.\end{equation}

The main advantage of this method is that its implementation requires only some vector matrix (or vector transposed matrix) products. 
It leads to a complexity in $\mathcal{O}(n^2),$ which may be reduced to $\mathcal{O}(n)$ if the matrix is sparse. 
Moreover, the scheme needs $\mathcal{O}(kn)$ operations, but the algorithm can be restarted  to minimize this contribution.

For each Newton step, we thus have to solve the linear problem $\bf{J}\delta\bf{X}=-\bf{F}$, which will be rewritten in this 
section as $\mathbf{A}\bf{x}=\mathbf{b}$ for simplicity.
Stationary iterative methods (Jacobi, Gauss-Seidel, SOR) approximate the unknown vector $\mathbf{x}$ with a scheme $\mathbf{x}^{(k)}=B\mathbf{x}^{(k-1)}+\mathbf{c,}$ where neither $B$ nor $\mathbf{c}$ depend upon the iteration count $k$. We use the  GMRES method, which is part of a class of (nonstationary) methods named Krylov subspace methods. The GMRES algorithm tries to evaluate the guess $\mathbf{x}_n \in \mathcal{K}_n$ (the $n$-th order Krylov subspace) by solving an optimization problem (minimizing the residue $\|\mathbf{A}\mathbf{x}_n-\mathbf{b}\|_{L_2}$) like a least-square problem.
The $n$-th Krylov subspace is spanned over the $n$ basis vectors,
\begin{equation}
\mathcal{K}_n = \text{span}(\mathbf{b}, \mathbf{A}\mathbf{b},\mathbf{A}^2\mathbf{b} . . ., \mathbf{A}^{n-1}\mathbf{b})
.\end{equation}
Then, the approximate solution is $\mathbf{x}_n=\mathbf{K}_n\mathbf{c}$, where the Krylov matrix is $\mathbf{K}_n = \left[\mathbf{b},\mathbf{A}\mathbf{b},\mathbf{A}^2\mathbf{b} . . ., \mathbf{A}^{n-1}\mathbf{b}\right]$ and $\mathbf{c}$ is an appropriate vector  such that
\begin{equation}
\min{\|\mathbf{A}\mathbf{x}_n-\mathbf{b}\|_{L_2}}=\min{\|\mathbf{A}\mathbf{K}_n\mathbf{c}-\mathbf{b}\|_{L_2}}
.\end{equation}
This vector is easily computed as it just requires a Matrix-vector product. This  is particularly well suited for sparse 
matrices because this operation evolves roughly as the number of nonzero elements.

A straightforward method to find the least-square solution of this problem would be to compute the $QR$ factorization of the matrix $\mathbf{A}\mathbf{K}_n$. Indeed, the normal equation $A^TAx=A^Tb\Leftrightarrow Rx=Q^Tb$ can be solved by back substitution. 
However, this is both unstable and too expensive (the QR factorization requires $2mn^2$ flops, $d=Q^Tb$, $2mn$ flops, and $Rx=d$, $n^2$ flops, where $m$ is  the iterate and $n$ is the order of the Krylov subspace). Moreover this natural Krylov basis is ill-conditioned and is thus not a good choice. Indeed, the basis vectors have to be linearly independent, and the successive multiplications by $A$ lead to vectors, which point (fastly) to the dominant eigenvector of $A$ and are thus numerically collinear. 
For this reason, we  build another space , $\mathcal{Q}_n$ , where the basis vectors are orthogonal (this orthogonalization of Krylov bases is named the Arnoldi method). 
During the Arnoldi step to build the orthogonal basis, the matrix $H$ of the orthogonalization elements 
is built. 
This matrix has a special shape (Hessenberg shape), which is close to triangular (one subdiagonal more).
Our minimization problem can be rewritten as
\begin{equation}
\min{\|\mathbf{A}\mathbf{x}_n-\mathbf{b}\|_{L_2}}=\min{\|\mathbf{A}\mathbf{Q}_n\mathbf{y}-\mathbf{b}\|_{L_2}}=\min{\|\mathbf{Q}_{n+1}\mathbf{H}_{n+1}\mathbf{y}-\mathbf{b}\|_{L_2}}
,\end{equation}
where $\mathbf{y}$ designs the new unknown ($\mathbf{Q}_n \mathbf{y} = \mathbf{x}_n$). 
\\

Furthermore, we have to combine this method with a global convergence strategy. Indeed, as one gets close to the solution, 
the choice of the descending direction may lead to spurious effects, especially if $\delta \textbf{X}$  becomes very large. 
Then,  to ensure the global convergence of the scheme, 
we use a combination of the line search and backtracking methods, which is an efficient method to solve 
nonlinear equations. Precisely, when $f_k > f_{k-1}$, $\delta \textbf{X}$ may be too large. 
Then, $\delta \textbf{X}$  is reduced by a factor $\lambda$, 
\begin{equation}
\textbf{X}_k=\textbf{X}_{k-1}+\lambda \delta \textbf{X}
,\end{equation}
with $\lambda$ chosen to minimize $f(\textbf{X}_{k-1}+\lambda \delta \textbf{X} )$ to maintain the iteration in the descending direction.

\subsection{Treatment of the singularity}

\begin{figure}\centering\includegraphics[width=\columnwidth]{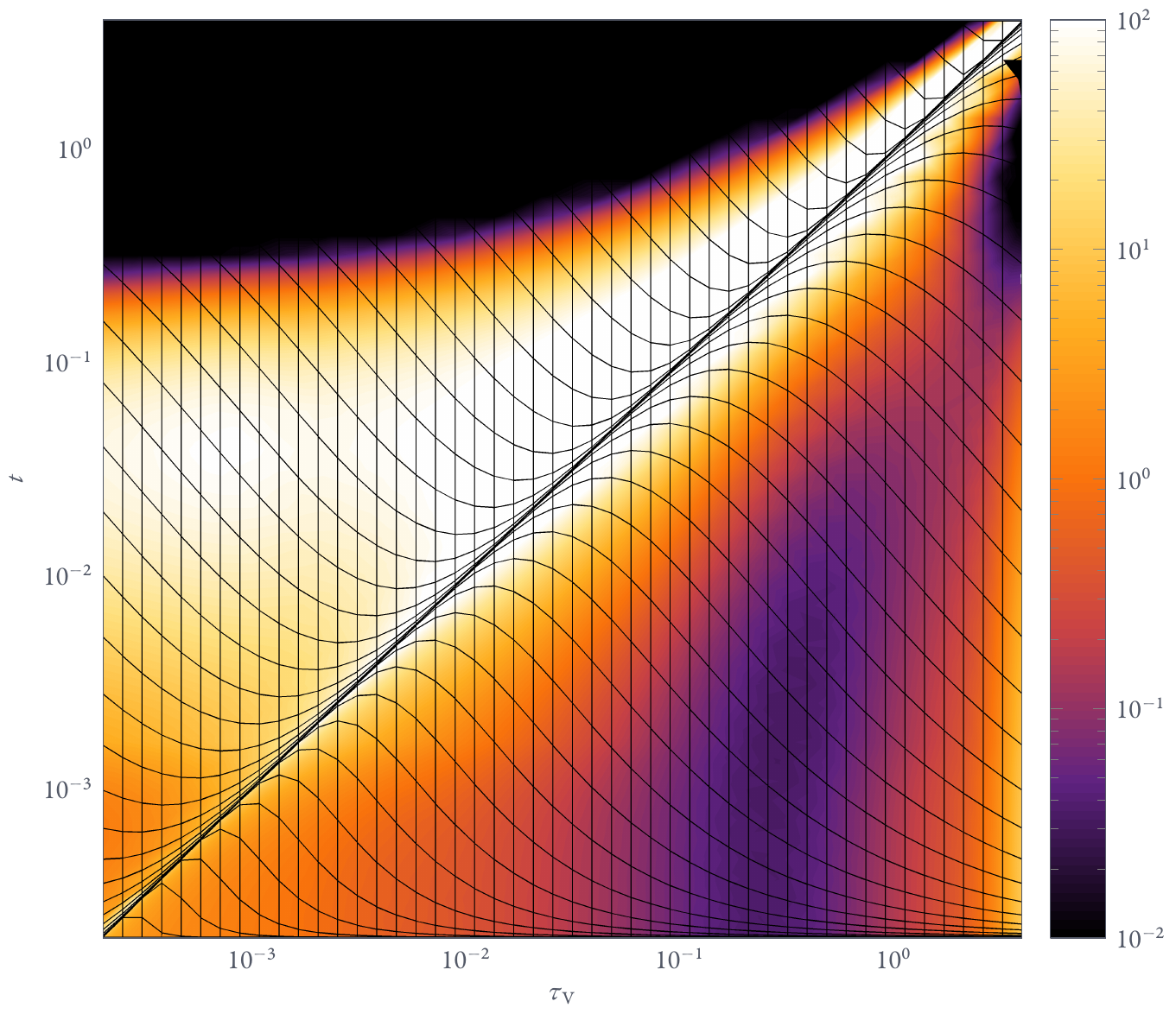}\caption{Map of the integrand of $\bar{\mathcal{J}}_{ij}$ for a radiative transition (see Eq. \ref{eq:Jtotc}). 
The diagonal is singular, and the quadrature in the periodization method produces a mesh refinement around it. The 
color encoding of the integrant is in logarithmic arbitrary units.}
  \label{fig:kernel}
\end{figure}

The computation of $\bar{\mathcal{J}}$ goes through the integration of a singular function. Indeed, in the integrand of 
Eq. (\ref{eq:Jtot}), the function $E_1(x)$ has a logarithmic divergence at $x=0^+$, see Fig \ref{fig:kernel}.

\begin{figure}\centering\includegraphics[width=\columnwidth]{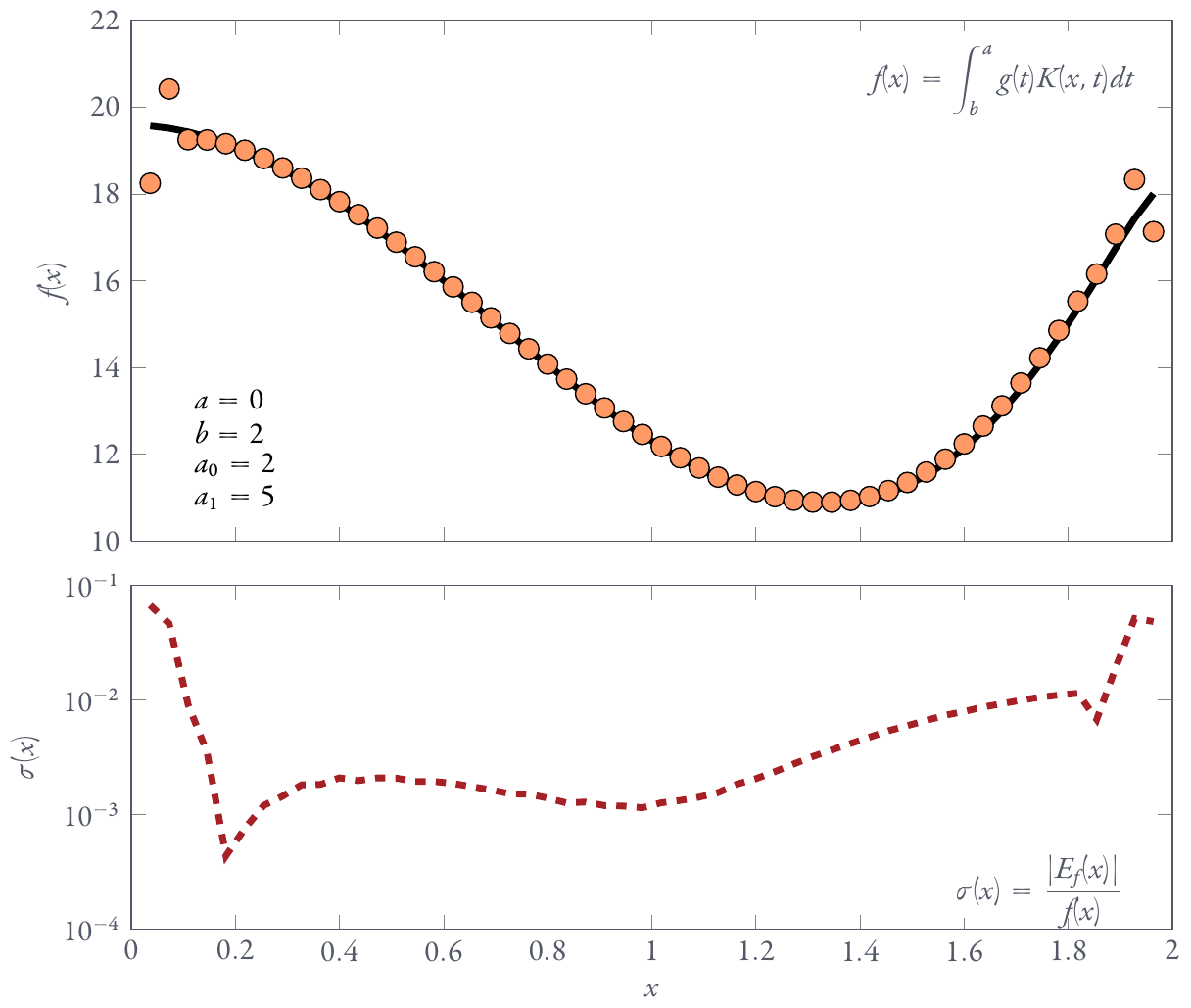}\caption{Benchmarking of the periodization method applied to the integration with a logarithmically singular kernel ($g(t) = a_0+a_1t  $, $K(x,t) = -\gamma - ln(|G(t)-G(x)|) $ and $G(x)=\int_0^t{g(s)ds}$). 
The integration domain has been split into two subdomains (see text for details).}
  \label{fig:benchmarkperio}
\end{figure}

To integrate this kernel, we use the periodization method developed by \cite{Helluy:1998}, which consists 
of a high order quadrature method with a variable change. The new variable is a polynomial of degree 
$k$, properties of which improve on the order of rectangle rule. The advantage of this method is that it can handle 
singular functions. Indeed the error 
$|E_N(f)|$ on an interval subdivided in $N$ subintervals goes as $\sim C_k / N^\gamma$, where $\gamma \rightarrow (2k - 1)$ if $f$ has a logarithmic 
singularity at one bound of the integration domain (see Table \ref{tab:err}).
\begin{table}[!ht]
\caption{Benchmark of the periodization method for a logarithmic singularity  (from \cite{Helluy:1998}). }
  \label{tab:err}
\begin{center}
\begin{tabular}{cccccc}
  \hline
  \hline
  $f(x)$        & $k$ & $C_k$ & $\gamma$ & $|E_{10}(f)|$ & $|E_{20}(f)|$ \\
  \hline
$\ln(x)$        & 2  & 3.22$\times$10$^{0}$     & 3.03 & 0.30$\times$10$^{-2}$ & 0.37$\times$10$^{-5}$\\
                        & 3  & 3.34$\times$10$^{2}$     & 5.05 & 0.31$\times$10$^{-3}$ & 0.93$\times$10$^{-5}$\\
                        & 5  & 3.05$\times$10$^{4}$     & 9.09 & 0.14$\times$10$^{-4}$ & 0.42$\times$10$^{-7}$\\
                        & 10 & 1.74$\times$10$^{14}$    & 19.1 & 0.48$\times$10$^{-5}$ & 0.92$\times$10$^{-11}$\\
  \hline
\end{tabular}
\end{center}
\end{table}
However the periodization method requires that the integration is done over the interval $[0,1]$ and 
that the singularity is located at one extremum of this interval. We thus split the integration domain 
into two subdomains, $[{\tau}^{in}_V,{\tau}_V]$ and $[{\tau}_V,{\tau}^{out}_V]$ , and then normalize 
each subdomain to $[0,1]$. However, since the MPI strategy limits the number of model layers 
handled by each process, this can lead to the undersampling of one of the subdomains, when ${\tau}_V$ 
approaches one of the bounds of the total domain (i.e., the boundaries of the model atmosphere). 
We have thus tested this method under the same conditions as those met in our problem, without any resampling 
of the subdomains. 
The result is displayed in Fig. \ref{fig:benchmarkperio}. As expected, the relative error strongly increases when 
one of the subdomains includes too few model layers (typically fewer than 4) and becomes unacceptable 
for the extreme points. They will thus be considered  ghost points in all subsequent computations. 
This does not affect the computation of the statistical equilibrium.
Because we can compute the nonlinear function $\textbf{F}(\textbf{X})$ with good accuracy, the next step to find its root is to compute the Jacobian matrix  to solve the system through a Newton's scheme.
%


\subsection{Computation of the Jacobian matrix}
The method presented here, and more generally the solving of nonlinear systems of equations, requires 
the computation of the Jacobian matrix. One of the advantages of the method we develop is that an analytical 
form can be derived. Here, the function $\textbf{F}(\textbf{X})$ is a set of $N \times P$ equations, $N$ being the 
number of energy levels of the species considered, and $P$ the number of layers in the discretized model 
atmosphere. The differentiation of $F_i^l$ over $x_k^m$ ($i$ and $k$ referring to energy levels, and $l$ and $m$ 
to atmosphere layers) leads to a set of integro-differential equations, with the following four cases:

\begin{subequations}
\begin{align}
 \text{if } k = i  &\text{ and } l=m \label{eq:jac1} \\ 
\frac{{\partial F_i^l}}
    {{\partial x_i^l}} =& \sum\limits_{j \ne i}^N\frac{\partial \bar{\mathcal{J}}^l_{ij}}{\partial x_i^l}
      \left( {x_i^l B_{ij}  - x_j^l B_{ji} } \right) \nonumber
   + \sum\limits_{j \ne i}^N{\left( A_{ji}+B_{ji}\bar{\mathcal{J}}_{ji}+n_{col} C_{ji} \right), } 
  \nonumber \\
      %
       \text{if } k \neq i &\text{ and } l=m \label{eq:jac2} \\
\frac{{\partial F_i^l}}
    {{\partial x_k^l}} =& \frac{\partial \bar{\mathcal{J}}^l_{ik}}{\partial x_k^l}
      \left( {x_k B_{ki}  - x_i B_{ik} } \right)
    - {\left( A_{ki}+B_{ki}\bar{\mathcal{J}}_{ki}+n_{col} C_{ki} \right), }
    \nonumber \\
    %
     \text{if } k = i  &\text{ and } l \neq m \label{eq:jac3} \\
    \frac{{\partial F_i^l}}
    {{\partial x_i^m}} =& \sum\limits_{j \ne i}^N \frac{\partial \bar{\mathcal{J}}^l_{ij}}{\partial x_i^m}
      \left( {x_i B_{ij}  - x_j B_{ji} } \right), \nonumber \\
    %
     \text{if } k \neq i  &\text{ and }  t \neq m \label{eq:jac4} \\
\frac{{\partial F_i^l}}
    {{\partial x_k^m}} =& \frac{\partial \bar{\mathcal{J}}^l_{ik}}{\partial x_k^m}
      \left( {x_k B_{ki}  - x_i B_{ik} } \right). \nonumber
\end{align}
\end{subequations}

Equations \eqref{eq:jac1} and \eqref{eq:jac2} represent local couplings.  Mathematically, they correspond 
to the diagonal and block-diagonal parts of the Jacobian, respectively. Physically,  Eq. \eqref{eq:jac1} 
describes perturbations of the statistical equilibrium of a given level, by modifying the population of this level, 
within one atmospheric layer. Equation \eqref{eq:jac2}  describes the perturbations of the statistical equilibrium 
of a given level, by modifying the population of other levels, still within one atmospheric layer. 
Equations \eqref{eq:jac3} and \eqref{eq:jac4} correspond to off-diagonal blocks and describe nonlocal perturbations, i.e.,
the influence of the populations in another layer. Naturally, in these last differentiations the coupling is purely radiative, 
whereas the first and second differentiations include collisional processes.

To determine the terms $\partial \bar{\mathcal{J}}^l_{ik}/\partial x_k^m$, one needs to differentiate an integral expression 
over the variable $x_k^m$. This derivation must be considered  a functional derivative, and be performed before discretization (see Appendix \ref{appB} for the demonstration), 

\begin{equation}
\frac{{\delta \bar{\mathcal{J}}_{ij} \left[x_k\right]\left( \tau _{\text{v}} \right)}}
      {{\delta x_k \left( s \right)}}=\lim\limits_{\varepsilon \to 0}\frac{\bar{\mathcal{J}}_{ij} \left[x_k(\tau_\mathrm{V}) + \varepsilon \delta(\tau_\mathrm{V} - s)\right]-\bar{\mathcal{J}}_{ij} \left[x_k(\tau_\mathrm{V})\right]}{\varepsilon} 
,\end{equation}

where $\delta(\tau_\mathrm{V} - s)$ is the Dirac distrbution. 

In the case $\tau_\mathrm{V} \neq s$ ($l \neq m$), the derivatives for each component are

\begin{subequations}
\begin{align}
\label{eq:jac11}
\begin{split}
  \frac{\delta \bar{\mathcal{J}}_{ij}^\mathrm{Line}[x_i,x_j](\tau_\mathrm{V}) }
  {\delta x_i(s)}
   &= D_{ij} \zeta(s) K(s,\tau_\mathrm{V}) + D_{ij} \zeta(s) E_{ij} \\
   &\times
    \begin{cases}
      \displaystyle
      \int_k^{\tau_\mathrm{V}^\mathrm{Out}}{\zeta(t) x_i(t) \tilde{K}_0(t,\tau_\mathrm{V},s)dt} & \text{if $k > \tau_\mathrm{V}$}\\
      \displaystyle
      \int^k_{\tau_\mathrm{V}^\mathrm{In}}{\zeta(t) x_i(t) \tilde{K}_0(t,\tau_\mathrm{V},s)dt} & \text{if $k < \tau_\mathrm{V}$}
    \end{cases},
  \end{split} \\
  \label{eq:jac22}
\begin{split}
  \frac{\delta \bar{\mathcal{J}}_{ij}^\mathrm{Line}[x_i,x_j](\tau_\mathrm{V}) }
  {\delta x_j(s)}
   &= -D_{ij} \zeta(s)
 E_{ij} \\
 &\times
    \begin{cases}
      \displaystyle
      \int_k^{\tau_\mathrm{V}^\mathrm{Out}}{\zeta(t) x_i(t) \tilde{K}_0(t,\tau_\mathrm{V},s)dt} & \text{if $k > \tau_\mathrm{V}$}\\
      \displaystyle
      \int^k_{\tau_\mathrm{V}^\mathrm{In}}{\zeta(t) x_i(t) \tilde{K}_0(t,\tau_\mathrm{V},s)dt} & \text{if $k < \tau_\mathrm{V}$}
    \end{cases},
    \end{split} \\
    \label{eq:jac33}
\begin{split}
  \frac{\delta \bar{\mathcal{J}}_{ij}^\mathrm{Cont}[x_i,x_j](\tau_\mathrm{V}) }
  {\delta x_i(s)}
   &= \zeta(s)
 E_{ij} \\
 &\times
    \begin{cases}
      \displaystyle
      \int_k^{\tau_\mathrm{V}^\mathrm{Out}}{\xi_{{ij}}(t)B_{{ij}}(t) \tilde{L}_0(t,\tau_\mathrm{V},s)dt} & \text{if $k > \tau_\mathrm{V}$}\\
      \displaystyle
      \int^k_{\tau_\mathrm{V}^\mathrm{In}}{\xi_{{ij}}(t)B_{{ij}}(t) \tilde{L}_0(t,\tau_\mathrm{V},s)dt} & \text{if $k < \tau_\mathrm{V}$}
    \end{cases},
   \end{split} \\
   \label{eq:jac44}
\begin{split}
  \frac{\delta \bar{\mathcal{J}}_{ij}^\mathrm{Cont}[x_i,x_j](\tau_\mathrm{V}) }
  {\delta x_j(s)}
   &= -\zeta(s)
 E_{ij} \\
 & \times
    \begin{cases}
      \displaystyle
      \int_k^{\tau_\mathrm{V}^\mathrm{Out}}{\xi_{{ij}}(t)B_{{ij}}(t) \tilde{L}_0(t,\tau_\mathrm{V},s)dt} & \text{if $k > \tau_\mathrm{V}$}\\
      \displaystyle
      \int^k_{\tau_\mathrm{V}^\mathrm{In}}{\xi_{{ij}}(t)B_{{ij}}(t) \tilde{L}_0(t,\tau_\mathrm{V},s)dt} & \text{if $k < \tau_\mathrm{V}$}
    \end{cases}.
       \end{split}
       \end{align}
\end{subequations}

However $\bar{\mathcal{J}}$ is not differentiable at $x(t)=x(\tau_\mathrm{V})$ because of the singularity, therefore
Eq. \eqref{eq:jac11} is undetermined for $s=\tau_\mathrm{V}$.
For these local terms, we thus need an approximate expression of the radiation field. We emphasize that we only apply this approximation to the computation of the Jacobian matrix. 

\begin{figure}\centering\includegraphics[width=\columnwidth]{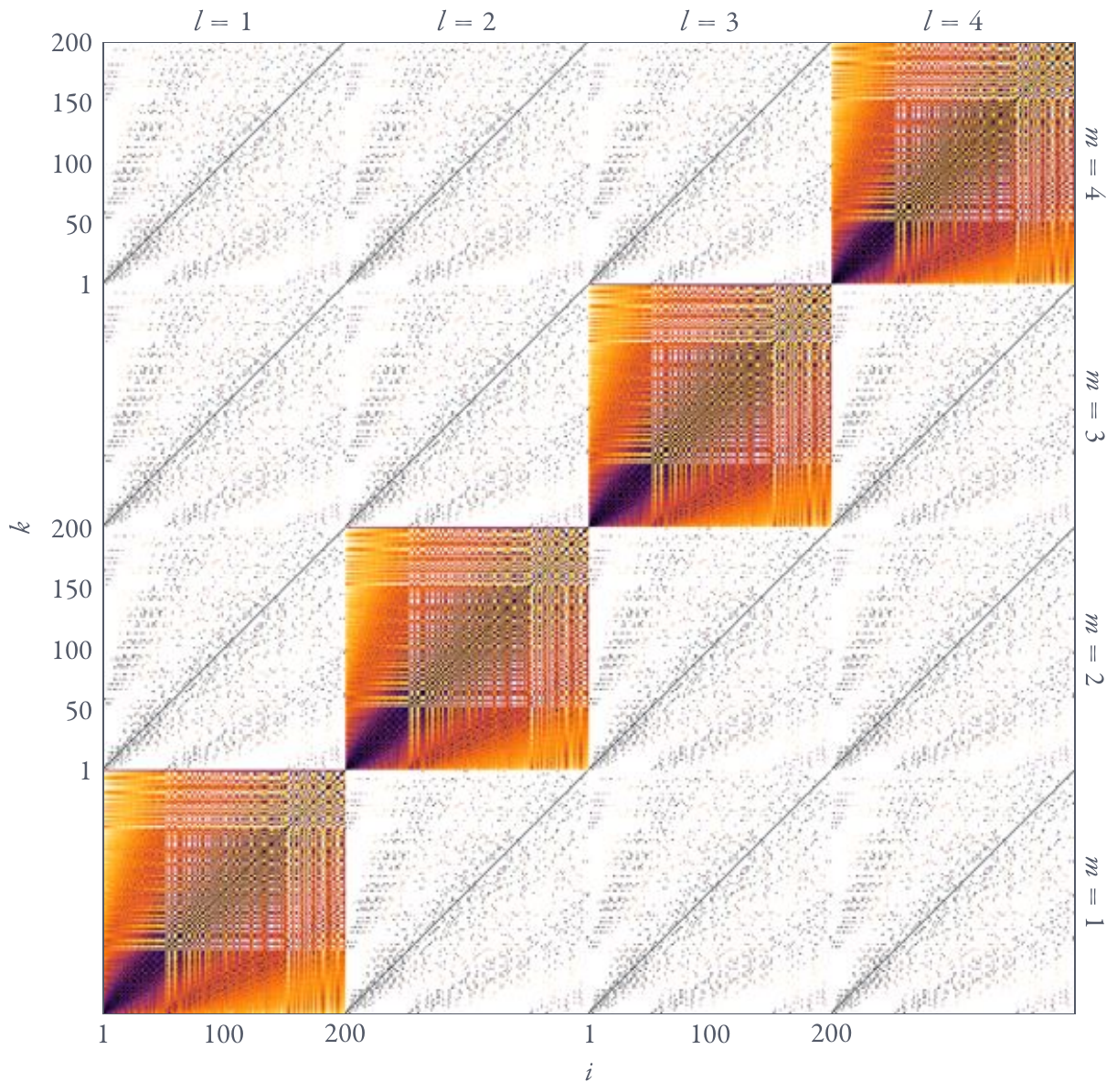}\caption{Pattern of the Jacobian matrix for 4 model atmosphere layers and 200 energy levels.}
  \label{fig:jacobian}
\end{figure}

\subsection{Sparsity of Jacobian matrix}
In Section \ref{solving}, we argue that GMRES is a well-adapted method for large-scale problems 
if the Jacobian matrix is sparse, 
as it only requires matrix-vector products, and no matrix inversion.  
The sparsity (or density) coefficient is defined by the ratio of nonzero elements over the total number of elements. The 
Jacobian matrix is composed by $P \times P$ submatrix of $N \times N $ elements leading to  $NP\times NP$ elements. The block diagonal of this matrix is full because of the irreducibility of the collisional coupling within a given layer (no selection rules for collisions). The number of nonzero elements in the block-diagonal part is $P \times N^2$. 
In the $P(P-1)$ off-diagonal submatrices, the nonzero elements are only due to radiative coupling between different layers, 
and their number is thus linked to the selection rules for allowed transitions. These rules thus lead to a sparsity of each submatrix with a density $\rho_\text{mol}=N_\text{transitions}/N^2$. The density of the complete Jacobian matrix is then given by
\begin{equation}
\rho=\frac{P\times N^2+P(P-1)\times N_\text{transitions}}{N^2P^2}=\frac{1+\rho_\text{mol}{(P-1)}}{P}
.\end{equation}
\begin{figure}[h!]
\centering
\includegraphics[width=\columnwidth]{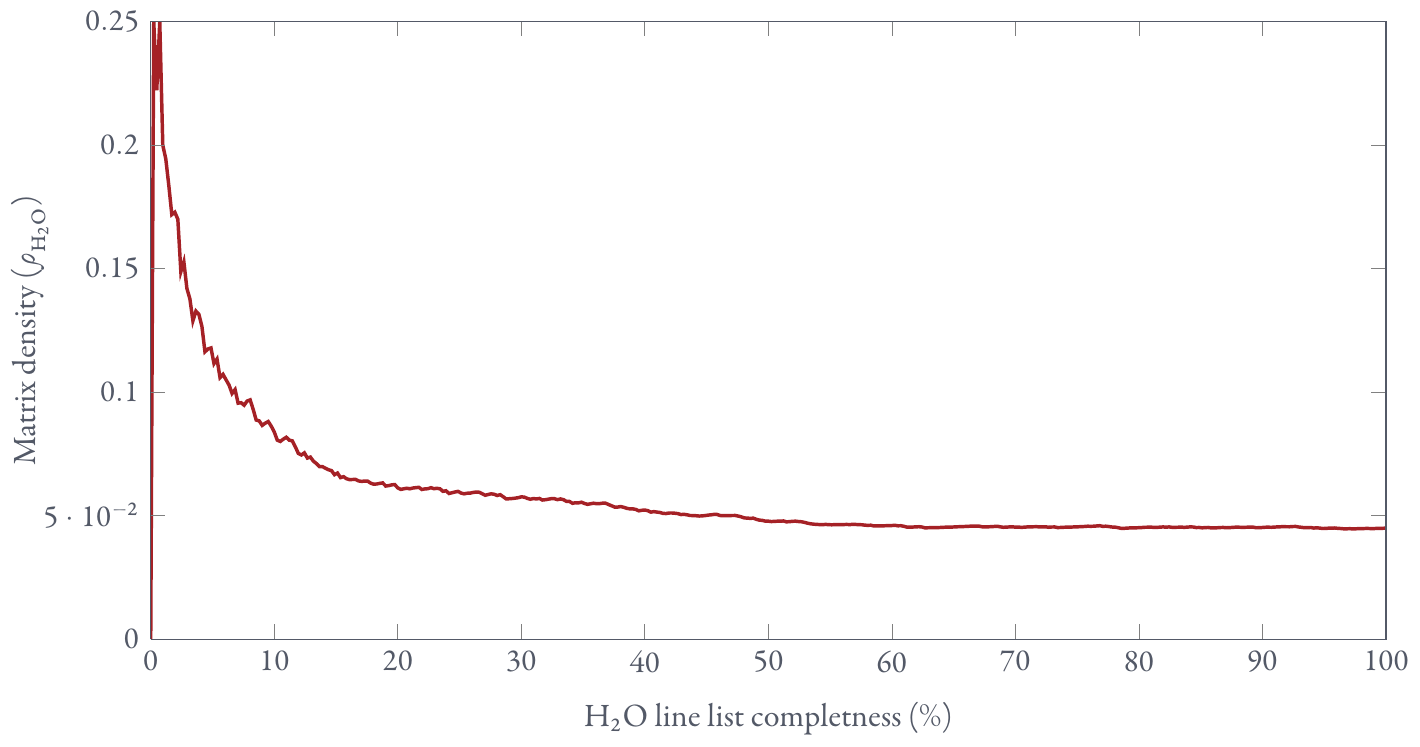}
\caption{Example of the evolution of the off-diagonal, block-matrix sparsity ($\rho_\text{mol}$) according to the number of considered levels. This example is computed for the 411 levels of the ortho--H$_2$O molecule. }
\label{fig:rhomol1}
\end{figure}
\begin{figure}[h!]
\centering
\includegraphics[width=\columnwidth]{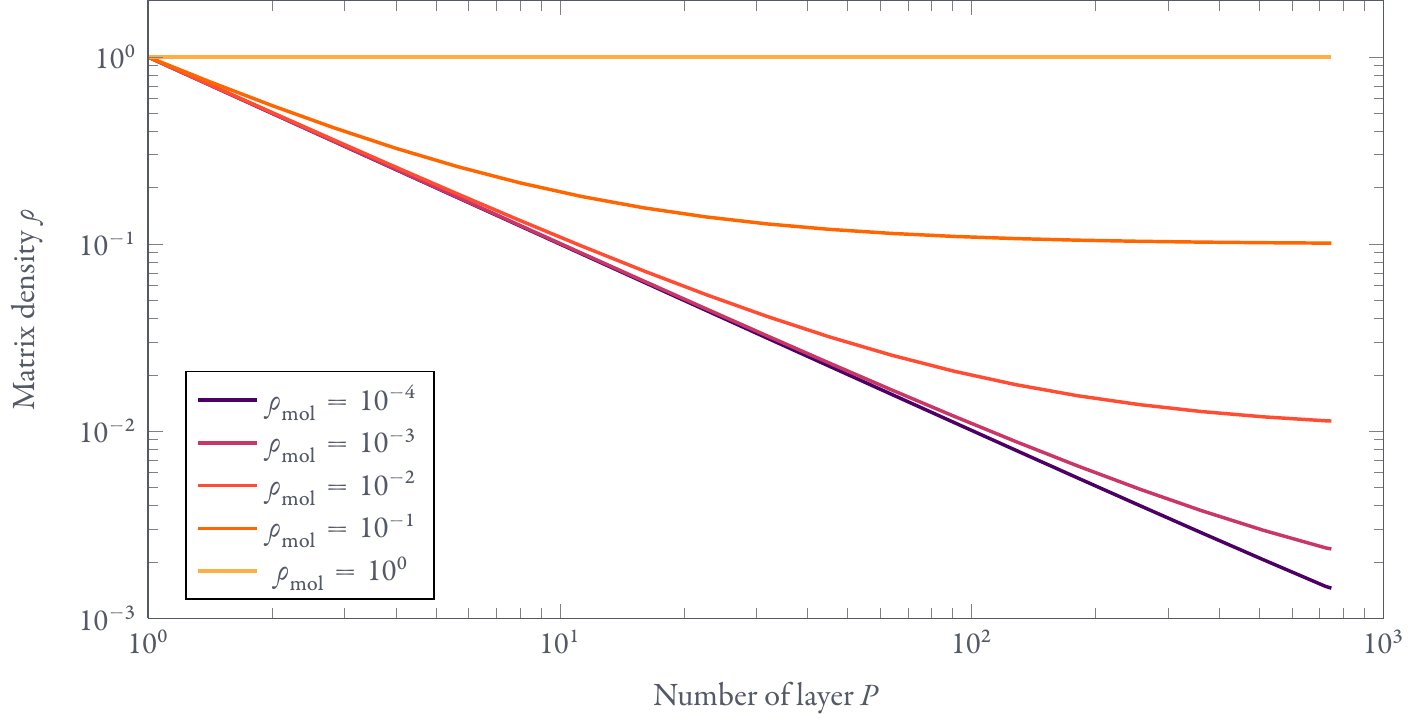}
\caption{Evolution of the Jacobian matrix sparsity according to the number of atmospheric layers for different values 
of $\rho_\text{mol}$. }
\label{fig:rhomol2}
\end{figure}
Then, as $\rho_\text{mol}(N)$ decreases rapidly with $N$ (e.g., $\rho_{\text{H}_2\text{O}}\sim 10^{-3}$,  for 
$N = 411$, see Figs. \ref{fig:rhomol1} and \ref{fig:rhomol2}) and $\rho$ is inversely proportional to $P$, 
the larger the system, the sparser 
the Jacobian matrix.  The GMRES scheme is thus well adapted and asymptotically approaches the $NP$ complexity.

\subsection{Jacobian block-diagonal approximation}

Assuming that $\tau_T(\tau_\mathrm{V})-\tau_T(t) \propto (t-\tau_\mathrm{V})$ (first order linearization) 
when $t$ is close to $\tau_\mathrm{V}$, 
because the kernel decreases when $|t-\tau_\mathrm{V}|$ increases, we assume that every function at $t$ is equal to its value 
at $\tau_\mathrm{V}$.

Splitting the integral over $t$ at $\tau_\mathrm{V}$ and inverting the integrals over $t$ and $\mu$,  \cite{Gonzalez-Garcia:2008}
show that the mean intensity can be rewritten as 
\begin{subequations}
\begin{align}
\bar{\mathcal{J}}_{ij}(\tau_\mathrm{V}) & =\frac{x_i(\tau_\mathrm{V})}{x_j(\tau_\mathrm{V})-x_i(\tau_\mathrm{V})}\left(1-\beta_-(\tau_\mathrm{V})-\beta_+(\tau_\mathrm{V})\right) \\
\bar{\mathcal{J}}_{ij}(\tau_\mathrm{V}) & = \tau_{max}I_\nu\beta(\tau_\mathrm{V}), 
\end{align}
\end{subequations}
where $\beta_-(\tau_\mathrm{V})$ and $\beta_+(\tau_\mathrm{V})$ are inward and outward escape 
probabilities (defined in  \cite{Gonzalez-Garcia:2008}). The $\beta(\tau_\mathrm{V})$ invoke the function $E_2(x),$ 
which is not singular and differentiable everywhere in the domain. This enables an analytic estimate of the bloc diagonal terms.
The pattern of the Jacobian is displayed in Fig. \ref{fig:jacobian}. 

\subsection{The numerical code MOrad}

The above strategy has been implemented in a numerical code named MOrad written in Fortran 90. It takes as inputs:
\begin{enumerate}
\item A model atmosphere (temperature, atomic, molecular, 
and electronic densities etc. as a function of geometrical thickness) 

\item Spectroscopic data (level energies, statistical weights, radiative rates)

\item Collisional rates 
\end{enumerate}
Then it computes the continuous opacities and uses an initial guess to calculate the function \eqref{eq:fonction} and its Jacobian matrix.
Choosing the initial guess $\textbf{X}_0$ is an important operation. In MOrad we tested the use of an LTE solution (Boltzmann distribution), 
and a 0 K solution (the molecules in the ground level). We find that the convergence rate is better with the "0 K" initial guess.

The singular integrations needed to compute $\bar{\mathcal{J}}$ are performed with the periodization method. 
The Green's function needed to compute $\bar{\mathcal{J}}$ is stored at each point. The required space 
scales as the number of layers times the number of frequency points (in a parallel code; otherwise it would scale with 
the square of the number of layers). 
The evaluation of the function and the Jacobian are parallelized with MPI. 
At first glance, two parallelization strategies may be considered: either a parallelization over frequencies 
or over atmosphere layers. However,  \cite{vanNoort2002} show that the most efficient parallelization to reduce
communications is  spatial parallelization because of the strong local coupling between energy levels. 
The populations of all energy levels within each atmosphere layer should thus be computed by the same processor, 
with several adjacent layers  possibly treated by the same processor. 
 In MOrad we adopt a subdomain decomposition where each processor computes the physical values needed in one given layer and solves the local statistical equilibrium. 
Integrations are performed with a scalable algorithm, which minimizes global communications  to preserve a good scalability. 

The code has been interfaced with a nonlinear solver of the parallel and scalable library PETSc \citep[see, e.g.,][]{petsc-web-page,petsc-user-ref,petsc-efficient}. This solver takes 
as input the distributed function vector and the Jacobian matrix and returns the root of the function  after preconditioning.

The main advantages of this library are its performance and its scalability. Moreover, PETSc provides access to many different nonlinear 
methods and preconditioning, which allows us a versatile choice of the best method to be adapted for a given physical problem. Other efficient libraries, such as the parallel IO library HDF5, are interfaced with MOrad  to preserve the performances for large-scale problems and the 
portability of output data. Future planned developments include an MPI/OpenMP hybridization (openMP parallelization over 
frequencies). 

\section{Discussion}

\subsection{Example of application}

\begin{figure}[h]
\centering
\includegraphics[width=\columnwidth]{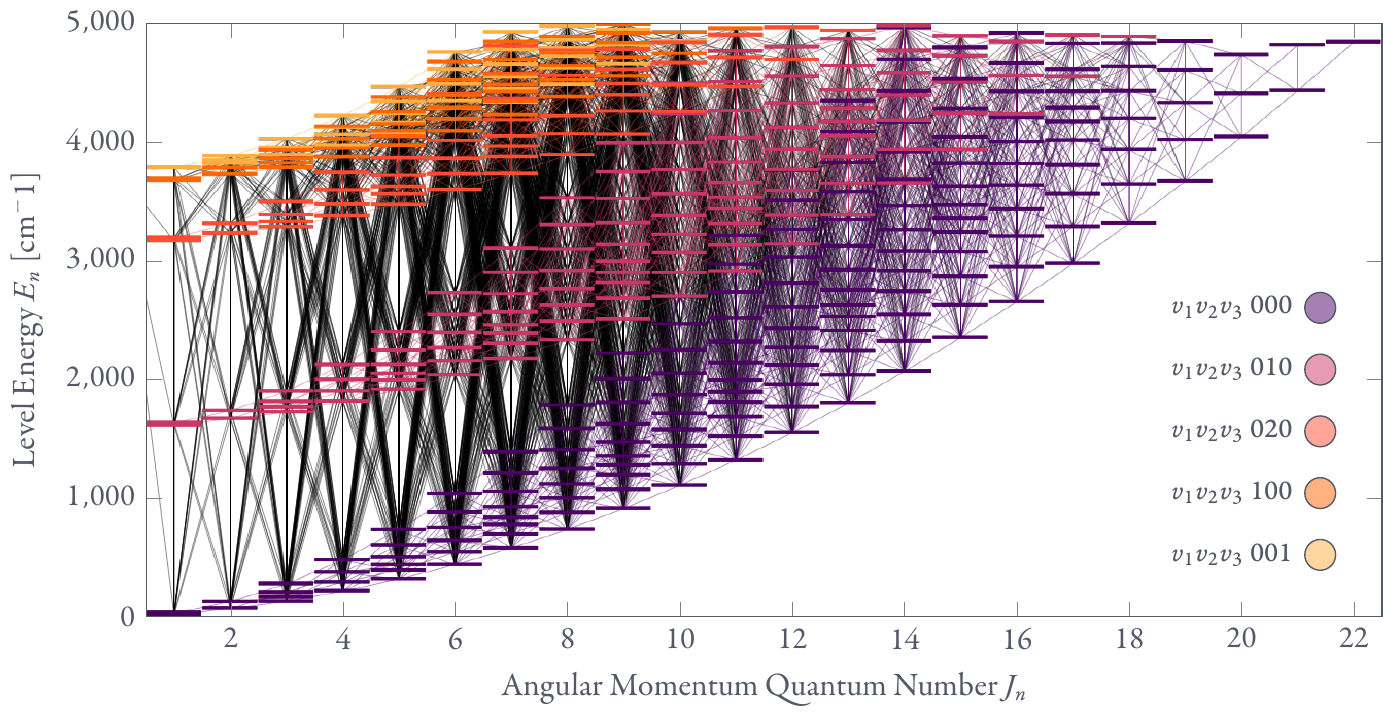}
\caption{Grotrian diagram of the ortho--H$_2$O molecule showing the radiative transitions taken into account. The parameter $J_n$ is the rotational angular momentum of H$_2$O. Colors indicate different vibrational states. }
\label{fig:godrian}
\end{figure}

To illustrate the applicability of MOrad, we compute the nonLTE departure coefficients of water molecules in a MARCS \citep{Gustafsson:2008} 
red supergiant model atmosphere. 
The model considered here has an effective temperature of 3500 K, 
a $\log g = 0$, a solar metallicity, and a microturbulence of 2 km~s$^{-1}$. These parameters correspond to typical red supergiant stars.
 We consider more than 800 rovibrational levels, 
leading to more than 330 000 transitions and 15 000 lines (see the Grotrian diagram of ortho H$_2$O in Fig. \ref{fig:godrian}). 
The energy levels and the radiative coefficients are taken from \cite{Barber2006}.
The collisional rates for H$_2$O--H$_2$ and H$_2$O--e$^-$ are taken from \cite{faure2008}.

\begin{figure}[h]
\centering
\includegraphics[width=\columnwidth]{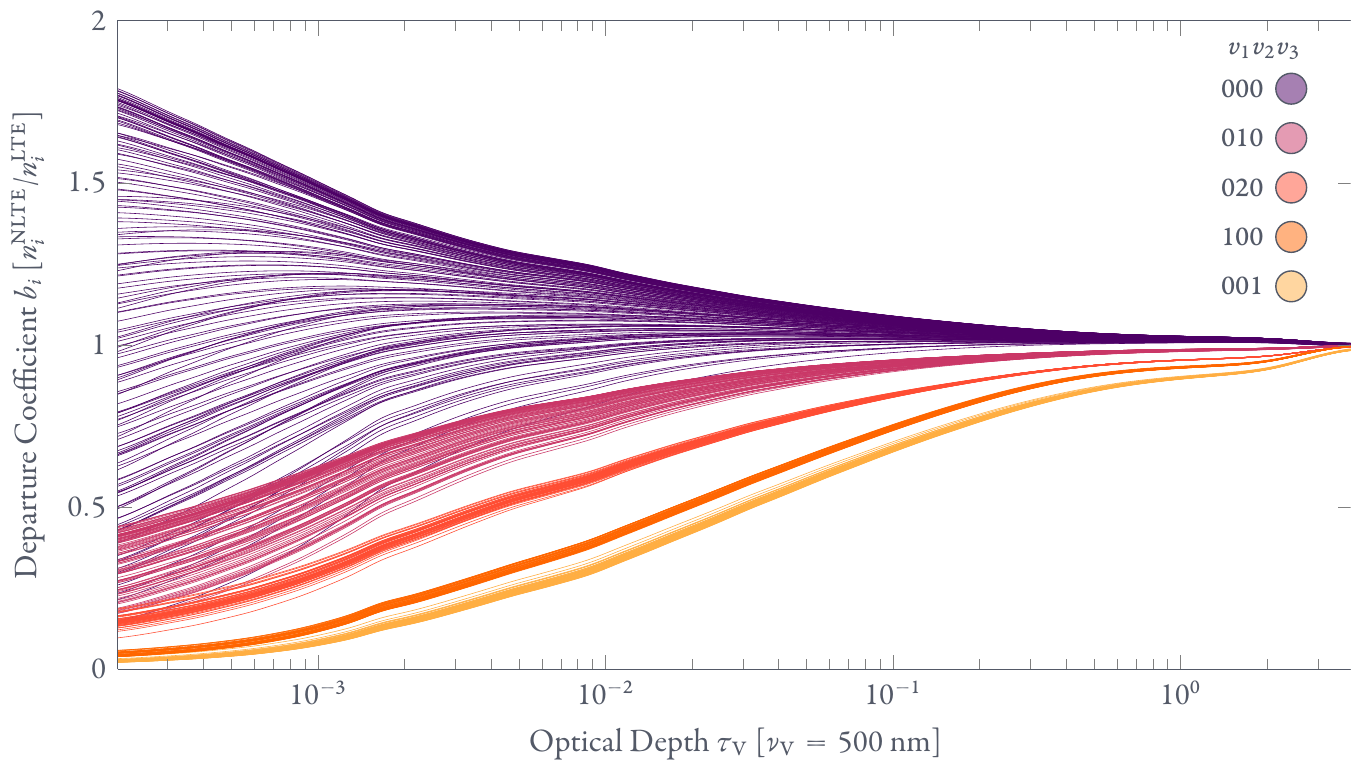}
\caption{LTE departure coefficients of ortho-H$_2$O in a red supergiant model atmosphere.}
\label{fig:deb}
\end{figure}

The code MOrad was launched on 47 processors using a GMRES Newton-Krylov method and a line-search global convergence strategy.
We  conducted preconditioning with the AMS method (Block Additive Schwarz method), where each subblock is preconditioned 
with an approximate iterative LU factorization method.
We obtain the root $\mathbf{X}^\star$ of the function ($\|\textbf{F}(\mathbf{X^\star})\|_\infty \leq 10^{-12} $) after three nonlinear 
iterations and a total time of code execution of less than $\sim$1h on the Alarik LUNARC system \citep[see][]{alarik}. The departure coefficients are presented in Fig. \ref{fig:deb}. 
The convergence rate is Q-quadratic, though a quadratic convergence rate would be expected. This may be because of  an over reconditioning or 
a  problem that is too simple(close to linear), but we obtain the same Q-quadratic convergence for a purely radiative problem.
\\

A detailed analysis of the results is out of the scope of this paper and is devoted to 
a forthcoming paper; a preliminary analysis was done in \cite{lambert2013}. In particular a detailed discussion 
of the use of the super-level approximation, which seems appropriate for the vibrationally excited states,  will be 
presented.  In short, the non-LTE calculations lead to stronger lines, especially rotational lines in the fundamental 
vibrational state around 2 $\mu$m, compared to LTE computation.

\subsection{Performance and complementarity of the method}
The method presented here is naturally accurate as it is equivalent to finding the root of a function and thus avoids any false convergence (assuming the solution is unique)\footnote{The solution of the Eddington 
problem (the two-level atom in the isothermal optically thick medium) is recovered at the 1st iteration with preconditioning, 
and after five iterations without preconditioning (relative error $<$ 0.005 for $\epsilon = 10^{-5}$ and $\tau_{max} = 1000$). This is expected because the inversion of the matrix by the Krylov subspace method is well known to work in this kind of  situation. }. Furthermore, we emphasize that because the integration is analytic, the method leads naturally to null 
residue.  Concerning the performance of the method, the performances of the method are difficult to assess. The convergence rate 
depends upon the choice of both the solver and the preconditioning. These choices are not unique and may vary 
according to the considered problem (species and model atmosphere). The memory demands scales with the dimensionality 
of the considered problem. 

The computation of the function and the Jacobian matrix leads to a scheme with an asymptotic complexity 
in $\mathcal{O}(Nb_\text{freq} \times (N \times P)^2)$. The complexity of the resolution scheme is more difficult to evaluate because the preconditioned Newton-Krylov method has a cost in $N_\text{freq} \times N^2 \times P^2$ 
($N_\text{freq}$ being the number of frequency points sampling the line profile, $N$ the number of 
energy levels, and $P$ the number of layers in the discretized atmosphere), but invokes another term that evolves as 
the square of the iteration index. This term is partially eliminated by restarting GMRES and is not really important in practice because the number of iterations is  small after appropriate preconditioning. 

Furthermore, the performance is better than $\mathcal{O}(n^2)$ and probably close to $\mathcal{O}(n)$. Indeed, for a sparse matrix, 
GMRES evolves in $\mathcal{O}(n)$. Because the only nonzero elements of the off-diagonal blocks of the Jacobian matrix are due 
to radiative coupling between layers, the sparsity of $J$ evolves as the ratio of the number of radiative transitions over $N^2$.
Taking a complexity in $Nb_\text{freq} \times N^2 \times P^2$ is then an superior limit, which underevaluates the method scalability 
in real cases.

As a comparison, the computation of the equivalent function of $\mathbf{F}(\mathbf{X})$ in MULTI has a complexity evolving in 
$N \times P$ because of the Scharmer operator, which performs no numerical integration to evaluate diffusion integrals 
but only one quadrature point. The system is solved with a Newton-Raphson scheme, which has a cost in $N^3 \times P^2$ when using 
the full matrix or $N^3 \times P$ when using the local operator (block-diagonal matrix; M. Carlsson, priv. comm.).
The theoretical asymptotic speed-up for MOrad compared to MULTI are summarized in Table \ref{tab:speedup}. 

\begin{table}[!ht]
  \label{tab:speedup}
\begin{center}
\begin{tabular}{lcc}
\hline\hline\noalign{\smallskip}
     & \multicolumn{2}{c}{$t_\text{MULTI}/t_\text{MOrad}$} \\
     & Full Operator & OAB (diagonal)\\
\noalign{\smallskip}
\hline\noalign{\smallskip}
   $\rho=1$ & $N$ & $N/P$ \\
   $\rho=0$ & $PN^2$ & $N^2$ \\    
\noalign{\smallskip}
\hline\noalign{\smallskip}    
\end{tabular}
\end{center}
\caption{Theoretical asymptotic speed-up for two extreme cases in terms of $\rho=\text{sparsity}(\mathbf{J}) (0 \leq \rho \leq 1$) for MOrad, 
compared to MULTI with either full or local operator (OAB stands for Olson, Auer, \& Buchler Operator). }
\end{table}

From Table \ref{tab:speedup}, one can see that our method is complementary
to existing methods, such as direct linearization. Indeed, in
some cases, with a large number of levels and a small number of layers, the
expected time of computation is shorter.

\section{Conclusion}
We have developed a new method to solve the RTE in non-LTE conditions. Rather than solving the RTE itself, we solve the statistical equilibrium equation system, which includes an analytical formal solution of the RTE. One of the main advantages of this choice is that the coefficients of the associated Jacobian matrix can be computed exactly. With this approach, a nonlinear solver can be used. In other words, this method avoids any linearization or stationary iterative schemes, which may converge very slowly or even falsely in some extreme cases (a very low non-LTE parameter $\epsilon$ and/or highly discretized media). We chose a method based on the GMRES nonlinear method. It has been implemented in a code that uses the PETSc library. We pay special attention   to the proper treatment of the singularity that arises in the integration of the mean radiation field. 

Furthermore, this method is parallelized using MPI, which makes it applicable to large-scale systems such as molecules and atoms in stellar atmospheres or interstellar medium, as it has an excellent scalability with the number of energy levels. Indeed, our code has been successfully applied to the modeling of water in the atmospheres of red supergiants. The results will be presented in a forthcoming paper. Further developments of the code include hybrid parallelization, with openMP parallelization over frequencies. We then plan to make the code public.

\begin{acknowledgements}
We are grateful to Pascal Azerad for fruitful discussions and Karin Ryde for proofreading the English in the manuscript. We acknowledge financial support from the Swedish Karl Tryggers Foundation under grants CTS 12:408 \& CTS 13:388 and its investment in science and astrophysics. 
Moreover, we thank the French National Agency for Research (ANR) through program number ANR-06-BLAN-0105, and from "Programme National de Physique Stellaire" (PNPS) of CNRS/INSU, France. 
N. R. is a Royal Swedish Academy of Sciences Research Fellow supported by a grant from the Knut and Alice Wallenberg Foundation. Funds from Kungl. Fysiografiska S\"{a}llskapet i Lund and support from the Swedish Research Council, VR are gratefully acknowledged.
\end{acknowledgements}

\bibliographystyle{aa}

\begin{thebibliography}{30}
\expandafter\ifx\csname natexlab\endcsname\relax\def\natexlab#1{#1}\fi

\bibitem[{{Ahues} {et~al.}(2002){Ahues}, {D'Almeida}, {Largillier}, {Titaud},
  \& {Vasconcelos}}]{Ahues2002}
{Ahues}, M., {D'Almeida}, F., {Largillier}, A., {Titaud}, O., \& {Vasconcelos},
  P. 2002, Journal of Computational and Applied Mathematics, 140, 13

\bibitem[{{Auer}(1991)}]{Auer1991}
{Auer}, L. 1991, in NATO ASIC Proc. 341: Stellar Atmospheres - Beyond Classical
  Models, ed. L.~{Crivellari}, I.~{Hubeny}, \& D.~G. {Hummer}, 9

\bibitem[{Balay {et~al.}(2013{\natexlab{a}})Balay, Brown, , Buschelman,
  Eijkhout, Gropp, Kaushik, Knepley, McInnes, Smith, \& Zhang}]{petsc-user-ref}
Balay, S., Brown, J., , {et~al.} 2013{\natexlab{a}}, {PETS}c Users Manual,
  Tech. Rep. ANL-95/11 - Revision 3.4, Argonne National Laboratory

\bibitem[{Balay {et~al.}(2013{\natexlab{b}})Balay, Brown, Buschelman, Gropp,
  Kaushik, Knepley, McInnes, Smith, \& Zhang}]{petsc-web-page}
Balay, S., Brown, J., Buschelman, K., {et~al.} 2013{\natexlab{b}}, {PETSc}
  {W}eb page, http://www.mcs.anl.gov/petsc

\bibitem[{Balay {et~al.}(1997)Balay, Gropp, McInnes, \&
  Smith}]{petsc-efficient}
Balay, S., Gropp, W.~D., McInnes, L.~C., \& Smith, B.~F. 1997, in Modern
  Software Tools in Scientific Computing, ed. E.~Arge, A.~M. Bruaset, \& H.~P.
  Langtangen (Birkh{\"{a}}user Press), 163--202

\bibitem[{{Barber} {et~al.}(2006){Barber}, {Tennyson}, {Harris}, \&
  {Tolchenov}}]{Barber2006}
{Barber}, R.~J., {Tennyson}, J., {Harris}, G.~J., \& {Tolchenov}, R.~N. 2006,
  \mnras, 368, 1087

\bibitem[{{Cannon}(1973)}]{Cannon1973}
{Cannon}, C.~J. 1973, \jqsrt, 13, 1011

\bibitem[{{Castor}(1970)}]{Castor1970}
{Castor}, J.~I. 1970, \mnras, 149, 111

\bibitem[{{Damgaard} {et~al.}(1992){Damgaard}, {Hjorth}, \&
  {Thejll}}]{Damgaard:1992}
{Damgaard}, P.~H., {Hjorth}, P.~G., \& {Thejll}, P.~A. 1992, \aap, 254, 422

\bibitem[{{Dickel} \& {Auer}(1994)}]{Dickel1994}
{Dickel}, H.~R. \& {Auer}, L.~H. 1994, \apj, 437, 222

\bibitem[{{Fabiani Bendicho} {et~al.}(1997){Fabiani Bendicho}, {Trujillo
  Bueno}, \& {Auer}}]{FabianiBendicho1997}
{Fabiani Bendicho}, P., {Trujillo Bueno}, J., \& {Auer}, L. 1997, \aap, 324,
  161

\bibitem[{{Faure} \& {Josselin}(2008)}]{faure2008}
{Faure}, A. \& {Josselin}, E. 2008, \aap, 492, 257

\bibitem[{{Gonzalez Garcia} {et~al.}(2008){Gonzalez Garcia}, {Le Bourlot}, {Le
  Petit}, \& {Roueff}}]{Gonzalez-Garcia:2008}
{Gonzalez Garcia}, M., {Le Bourlot}, J., {Le Petit}, F., \& {Roueff}, E. 2008,
  \aap, 485, 127

\bibitem[{{Gustafsson} {et~al.}(2008){Gustafsson}, {Edvardsson}, {Eriksson},
  {J{\o}rgensen}, {Nordlund}, \& {Plez}}]{Gustafsson:2008}
{Gustafsson}, B., {Edvardsson}, B., {Eriksson}, K., {et~al.} 2008, \aap, 486,
  951

\bibitem[{{Hauschildt} {et~al.}(1995){Hauschildt}, {Starrfield}, {Shore},
  {Allard}, \& {Baron}}]{Hauschildt1995}
{Hauschildt}, P.~H., {Starrfield}, S., {Shore}, S.~N., {Allard}, F., \&
  {Baron}, E. 1995, \apj, 447, 829

\bibitem[{{Helluy} {et~al.}(1998){Helluy}, {Maire}, \& {Ravel}}]{Helluy:1998}
{Helluy}, P., {Maire}, S., \& {Ravel}, P. 1998, Academie des Sciences Paris
  Comptes Rendus Serie Sciences Mathematiques, 327, 843

\bibitem[{{Hubeny} \& {Mihalas}(2014)}]{Hubeny2014}
{Hubeny}, I. \& {Mihalas}, D. 2014, {Theory of Stellar Atmospheres}

\bibitem[{{Juvela}(2005)}]{Juvela2005}
{Juvela}, M. 2005, \aap, 440, 531

\bibitem[{{Klein} {et~al.}(1989){Klein}, {Castor}, {Dykema}, {Greenbaum}, \&
  {Taylor}}]{Klein1989}
{Klein}, R.~I., {Castor}, J.~I., {Dykema}, P.~G., {Greenbaum}, A., \& {Taylor},
  D. 1989, \jqsrt, 41, 199

\bibitem[{{Lambert} {et~al.}(2013){Lambert}, {Josselin}, {Ryde}, \&
  {Faure}}]{lambert2013}
{Lambert}, J., {Josselin}, E., {Ryde}, N., \& {Faure}, A. 2013, in EAS
  Publications Series, Vol.~60, EAS Publications Series, ed. P.~{Kervella},
  T.~{Le Bertre}, \& G.~{Perrin}, 111--119

\bibitem[{Lunarc(2013)}]{alarik}
Lunarc. 2013, Alarik system details,
  http://www.lunarc.lu.se/Systems/AlarikDetails

\bibitem[{{Ng}(1974)}]{Ng1974}
{Ng}, K.-C. 1974, \jcp, 61, 2680

\bibitem[{{Olson} {et~al.}(1986){Olson}, {Auer}, \& {Buchler}}]{Olson1986}
{Olson}, G.~L., {Auer}, L.~H., \& {Buchler}, J.~R. 1986, \jqsrt, 35, 431

\bibitem[{{Paletou} \& {Anterrieu}(2009)}]{Paletou2009}
{Paletou}, F. \& {Anterrieu}, E. 2009, \aap, 507, 1815

\bibitem[{{Rybicki}(1997)}]{Rybicki:1997}
{Rybicki}, G.~B. 1997, \apj, 479, 357

\bibitem[{{Scharmer} \& {Carlsson}(1985)}]{MULTI1985}
{Scharmer}, G.~B. \& {Carlsson}, M. 1985, Journal of Computational Physics, 59,
  56

\bibitem[{{Socas-Navarro} \& {Trujillo Bueno}(1997)}]{Socas1997}
{Socas-Navarro}, H. \& {Trujillo Bueno}, J. 1997, \apj, 490, 383

\bibitem[{{Trujillo Bueno} \& {Fabiani Bendicho}(1995)}]{Trujillo-Bueno:1995}
{Trujillo Bueno}, J. \& {Fabiani Bendicho}, P. 1995, \apj, 455, 646

\bibitem[{{{\v S}t{\v e}p{\'a}n} \& {Trujillo Bueno}(2013)}]{Stepan2013}
{{\v S}t{\v e}p{\'a}n}, J. \& {Trujillo Bueno}, J. 2013, \aap, 557, A143

\bibitem[{{van Noort} {et~al.}(2002){van Noort}, {Hubeny}, \&
  {Lanz}}]{vanNoort2002}
{van Noort}, M., {Hubeny}, I., \& {Lanz}, T. 2002, \apj, 568, 1066

\end{thebibliography}

\begin{appendix}

\onecolumn

\section{Formulation}
\label{appA}
One starts with the classical formulation of the radiative transfer equation in a plane parallel geometry,
\begin{equation}
\mu \frac{dI_\nu}{ds}=-\left(\chi_\nu^\text{Cont}+\chi_\nu^\text{Line}\right)I_\nu+
\eta_\nu^\text{Cont}+\eta_\nu^\text{Line}
.\end{equation}
The monochromatic extinction and emissivity coefficients are expressed with an explicit dependence on levels populations as follows:

\begin{subequations}%
  \begin{align}
    \label{eq:def1a}
    \chi_\nu^\mathrm{Line} & = \phi^{ul}_\nu  \chi_{ul} = \phi^{ul}_\nu  \frac{h{\nu_{ul}}}{4\pi}  (B_{lu}n_{l}-B_{ul}n_{u})\\
    \label{eq:def1b}
    \eta_\nu^\mathrm{Line} & = \phi^{ul}_\nu \eta_{ul} =  \phi^{ul}_\nu \frac{h{\nu_{ul}}}{4\pi}A_{ul}n_{u.}
  \end{align}%
\end{subequations}
By inserting relative populations $f_i=n_i/n^{\rm{k}}$ ,where $n^{\rm{k}}$ is the total density of the species labeled by $\rm{k}$, one gets
\begin{subequations}%
  \begin{align}
    \label{eq:def1a}
    \chi_\nu^\mathrm{Line} &  = \phi^{ul}_\nu  \frac{h{\nu_{ul }}}{4\pi}  n^{\rm{k}}(B_{lu}f_{l}-B_{ul}f_{u})\\
    \label{eq:def1b}
    \eta_\nu^\mathrm{Line} & =  \phi^{ul}_\nu \frac{h{\nu_{ul}}}{4\pi}  n^{\rm{k}} A_{ul}f_{u.}
  \end{align}%
\end{subequations}
One brings up the new variable $x_i=f_i/g_i$, and using the relation $B_{ul}g_u=B_{lu}g_l$, where $g_i$ is the statistical weight of the level $i$, yields

\begin{subequations}%
  \begin{align}
    \label{eq:def1a}
    \chi_\nu^\mathrm{Line} &  = \phi^{ul}_\nu  \frac{h{\nu_{ul}}}{4\pi}  n^{\rm{k}}g_{u}(B_{lu}\frac{f_{l}}{g_{u}}-B_{ul}\frac{f_{u}}{g_{u}})
    = \phi^{ul}_\nu  \frac{h{\nu_{ul}}}{4\pi}  n^{\rm{k}}g_{u}B_{ul}(x_{l}-x_{u}) \\
    \label{eq:def1b}
    \eta_\nu^\mathrm{Line} & =  \phi^{ul}_\nu \frac{h{\nu_{ul}}}{4\pi}  n^{\rm{k}} g_{u} A_{ul}\frac{f_{u}}{g_{u}} = \phi^{ul}_\nu \frac{h{\nu_{ul}}}{4\pi}  n^{\rm{k}} g_{u} A_{ul}x_{u.}
  \end{align}%
\end{subequations}
Including the formulation of $\chi_\nu^\text{Line}$ and $\eta_\nu^\text{Line}$ in the radiative transfer equation, one obtains
\begin{equation}
\mu \frac{dI_\nu}{ds} =-\left(\chi_\nu^\text{Cont}+\phi^{ul}_\nu  \frac{h{\nu_{ul}}}{4\pi}  n^{\rm{k}}g_{u}B_{ul}(x_{l}-x_{u})\right)I_\nu + \eta_\nu^\text{Cont}+\phi^{ul}_\nu \frac{h{\nu_{ul}}}{4\pi}  n^{\rm{k}} g_{u} A_{ul}x_{u}
.\end{equation}
In order to create a transition independent scale, the previous equation is divided by an arbitrarily chosen continuum 
extinction coefficient $\chi_{V}$, here $\chi_{V} \equiv \chi_{\nu=500nm}$. Moreover we express it per hydrogen atom to bring out the ratio $n^{\rm k}/n^{\rm H}$ for numerical reasons. This normalization is symbolized by a tilde ($\sim$) for other variables,

\begin{equation}
\mu \frac{dI_\nu}{ \chi_{V} ds}=- \left(\frac{\tilde{\chi}_\nu^\text{Cont}}{\tilde{\chi}_{V}}\frac{n^{\rm H}}{n^{\rm H}}+
\frac{h\nu_{ul}}{4\pi}\frac{n^{\rm{k}}}{n^{\rm H}} \frac{1}{\tilde{\chi}_{V}} B_{ul}g_u \left(x_l-x_u\right)\phi^{ul}_\nu \right) I_\nu
+ \frac{h\nu_{ul}}{4\pi} \frac{n^{\rm{k}}}{n^{\rm H}} \frac{1}{\tilde{\chi}_{V}} A_{ul}g_ux_u \phi^{ul}_\nu
+ \frac{\eta_\nu^\text{Cont}}{n^{\rm H} \tilde{\chi}_{V}} \frac{n^{\rm H} \tilde{\chi}_\nu^\text{Cont}}{\chi_\nu^\text{Cont}}
.\end{equation}
With $\xi_\nu=\frac{\tilde{\chi}_\nu}{\tilde{\chi}_{V}}$, $E_{ij}=\frac{h\nu_{ij}}{4\pi}B_{ij}g_i$, $\zeta=\frac{n^{\rm{k}}}{n^{\rm H}} \frac{1}{\tilde{\chi}_{V}}$, $D_{ij}=\frac{h\nu_{ij}}{4\pi}A_{ij}g_i$, $S_\nu=\frac{\eta_\nu^\text{Cont}}{\chi_\nu^\text{Cont}}$, and $d\tau_{V} = - \chi_{V} ds$. Assuming that the continuum is formed in LTE ($S_\nu^\text{Cont}=B_\nu$), one obtains
\begin{equation}
\mu\frac{dI_\nu}{d\tau_{V}}=\left[ \xi_\nu + E_{ul}\zeta\phi^{ul}_\nu(x_l-x_u)\right]I_\nu
- D_{ul}x_u\zeta\phi^{ul}_\nu-\xi_\nu B_\nu
.\end{equation}

\section{Jacobian matrix coefficients}
\label{appB}

We start with the analytic formulation of the mean intensity field for the line component (similar demonstration for contunium), i.e., 
\begin{equation}
  \bar{\mathcal{J}_{ij}}[x_i,x_j](\tau_\mathrm{V}) = 
  D_{ij}\int_{\tau_\mathrm{V}^\mathrm{In}}^{\tau_\mathrm{V}^\mathrm{Out}}
  {
    \zeta(t) x_i(t) K[x_i,x_j](t,\tau_\mathrm{V}) dt
  }
.\end{equation}
Because the functional derivative of a functional product is
\begin{equation}
  \frac{\delta (F[y]G[y]) }{\delta y(x)} = G[y]\frac{\delta F[y] }{\delta  y(x)} + F[y]\frac{\delta G[y] }{\delta y(x)}
,\end{equation}
we functionally derive the product $x_i(t) K[x_i,x_j]$ over $x_i(\tau_\mathrm{V}=s)$, which yields,
\begin{equation}
  \label{eq:DJsplit}
    \frac{\delta \bar{\mathcal{J}_{ij}}[x_i,x_j](\tau_\mathrm{V}) }
    {\delta x_i(s)}
     = D_{ij} \int_{\tau_\mathrm{V}^\mathrm{In}}^{\tau_\mathrm{V}^\mathrm{Out}}
    {\zeta(t) \frac{\delta x_i(t)}{\delta x_i(s)} K[x_i,x_j](t,\tau_\mathrm{V}) dt}
      + D_{ij} \int_{\tau_\mathrm{V}^\mathrm{In}}^{\tau_\mathrm{V}^\mathrm{Out}}
    {\zeta(t) x_i(t) \frac{\delta K[x_i,x_j](t,\tau_\mathrm{V})}
      {\delta x_i(s)} dt}
.\end{equation}
Equation (\ref{eq:DJsplit}) can be rewritten as a sum of two integrals over $t$, $I_1$ and $I_2$.\\
The first term $I_1$ simplifies
\begin{equation}
  \begin{split}
    I_1 & = D_{ij} \int_{\tau_\mathrm{V}^\mathrm{In}}^{\tau_\mathrm{V}^\mathrm{Out}}
    {\zeta(t) \frac{\delta x_i(t)}{\delta x_i(s)} K[x_i,x_j](t,\tau_\mathrm{V}) dt}
     =  D_{ij} \int_{\tau_\mathrm{V}^\mathrm{In}}^{\tau_\mathrm{V}^\mathrm{Out}}
    {\zeta(t) \delta{(t - s)} K[x_i,x_j](t,\tau_\mathrm{V}) dt} \\
    & =  D_{ij} \zeta(s) K(s,\tau_\mathrm{V})
  \end{split}
.\end{equation}

This term is an indefinite form if $s=\tau_\mathrm{V}$, because the kernel $K(x,y)$ is singular at $x=y$.

We consider here the case $s \neq \tau_\mathrm{V}$ (the case $s  = \tau_\mathrm{V}$ is detailed in Section 3.4). 

\noindent The second term $I_2$ requires the computation of the functional derivative of the kernel,  which is
\begin{equation}
  K[x_i,x_j](t,\tau_\mathrm{V})
  = \frac{1}{2} \int_0^\infty
  {\phi_\nu(\tau_\mathrm{V}) \phi_\nu(t) E_1\left(\Delta \tau^\mathrm{Tot}_\nu [x_i,x_j](t,\tau_\mathrm{V}) \right) d\nu}
.\end{equation}
With the chain rule theorem of functional derivatives, we get
\begin{equation}
  \frac{\delta f(F[y]) }{\delta y(x)} = \frac{d f(F[y]) }{d F[y] } \frac{\delta F[y]}{\delta y(x)}
.\end{equation}
We then derive the kernel $K$,  finding
\begin{equation}
\label{eq:Dkern}
\begin{split}
\frac{\delta K[x_i,x_j](t,\tau_\mathrm{V})}{\delta x_i(s)} & = \frac{1}{2} \int_0^\infty
    \phi_\nu(\tau_\mathrm{V}) \phi_\nu(t) \times \frac{ \delta E_1\left(\Delta \tau^\mathrm{Tot}_\nu [x_i,x_j](t,\tau_\mathrm{V}) \right)}
      {\delta x_i(s) } d\nu \\
 & = -\frac{1}{2} \int_0^\infty \phi_\nu(\tau_\mathrm{V}) \phi_\nu(t)
\times E_0\left(\Delta \tau^\mathrm{Tot}_\nu [x_i,x_j](t,\tau_\mathrm{V}) \right) \frac{ \delta \Delta \tau^\mathrm{Tot}_\nu [x_i,x_j](t,\tau_\mathrm{V})}
      {\delta x_i(s) } d\nu
\end{split}
.\end{equation}
To simplify the calculus, we rewrite $\Delta \tau^\mathrm{Tot}_\nu$ as
\begin{equation}
  \label{eq:Deltatau}
  \begin{split}
    \Delta \tau^\mathrm{Tot}_\nu [x_i,x_j](t,\tau_\mathrm{V})
    & = \left| \int_t^{\tau_\mathrm{V}}{\xi_\nu(s') + E_{ij}\zeta(s')\phi_\nu(s') \left[x_j(s')-x_i(s')\right] ds'}\right| \\
    & = \int_{\tau_\mathrm{V}^\mathrm{In}}^{\tau_\mathrm{V}^\mathrm{Out}}
    \Theta(t,\tau_\mathrm{V},s')\left[\xi_\nu(s') + E_{ij}\zeta(s')\phi_\nu(s') \left[x_j(s')-x_i(s')\right]\right]ds'
  \end{split}
,\end{equation}
where the distribution $\Theta(t,\tau_\mathrm{V},s)=-2(\theta(t-\tau_\mathrm{V})-0.5) (\theta(s-t)-\theta(s-\tau_\mathrm{V}))$, and $\theta(x)$ is the Heaviside function. This distribution can be rewritten,
\begin{equation}
\Theta \left( {t,\tau _V,s } \right) =
\begin{cases}
 \Pi_{a,s}(t) & \text{if } s<\tau_\mathrm{V} \\
 \Pi_{s,b}(t) & \text{if } s>\tau_\mathrm{V}
\end{cases}
.\end{equation}
 \\
We functionally derive Eq. (\ref{eq:Deltatau}), yielding
\begin{equation}
  \begin{split}
    \frac{ \delta \Delta \tau^\mathrm{Tot}_\nu [x_i,x_j](t,\tau_\mathrm{V})}
    {\delta x_i(s) } 
    & = - \int_{\tau_\mathrm{V}^\mathrm{In}}^{\tau_\mathrm{V}^\mathrm{Out}}
    {\Theta(t,\tau_\mathrm{V},s')E_{ij} \zeta(s') \phi_\nu(s') \frac{\delta x_i(s')}{\delta x_i(s) }ds'} = - \int_{\tau_\mathrm{V}^\mathrm{In}}^{\tau_\mathrm{V}^\mathrm{Out}}
    {\Theta(t,\tau_\mathrm{V},s')E_{ij} \zeta(s') \phi_\nu(s') \delta(s - s') ds'} \\
    & = - E_{ij} \zeta(s) \Theta(t,\tau_\mathrm{V},s) \phi_\nu(s).
  \end{split}
\end{equation}
We insert this expression in Eq. (\ref{eq:Dkern}) and we define the modified kernel $\tilde{K}_0(t,\tau_\mathrm{V},s)$,
\begin{equation}
  \begin{split}
    \frac{\delta K[x_i,x_j](t,\tau_\mathrm{V})}{\delta x_i(s)}
    & = E_{ij} \zeta(s) \Theta(t,\tau_\mathrm{V},s) \frac{1}{2} \int_0^\infty
    {\phi_\nu(\tau_\mathrm{V}) \phi_\nu(t) \phi_\nu(s) E_0\left(\Delta \tau^\mathrm{Tot}_\nu [x_i,x_j](t,\tau_\mathrm{V}) \right) d\nu}\\
    & = E_{ij} \zeta(s) \Theta(t,\tau_\mathrm{V},s) \tilde{K}_0[x_i,x_j](t,\tau_\mathrm{V},s).
  \end{split}
\end{equation}
The integral $I_2$ from equation (\ref{eq:DJsplit}) can be rewritten as
\begin{equation}
  \begin{split}
    I_2 & = D_{ij} \int_{\tau_\mathrm{V}^\mathrm{In}}^{\tau_\mathrm{V}^\mathrm{Out}}
    {\zeta(t) x_i(t) \frac{\delta K[x_i,x_j](t,\tau_\mathrm{V})}
      {\delta x_i(s)} dt} \\
    & = D_{ij} E_{ij} \zeta(s) \int_{\tau_\mathrm{V}^\mathrm{In}}^{\tau_\mathrm{V}^\mathrm{Out}}
    {\Theta(t,\tau_\mathrm{V},s) \zeta(t) x_i(t) \tilde{K}_0[x_i,x_j](t,\tau_\mathrm{V},s) dt}
  \end{split}
,\end{equation}
which, according to the $\Theta$ properties, becomes
\begin{equation}
  I_2= D_{ij} E_{ij} \zeta(s) \times
  \begin{cases}
    \displaystyle
    \int_s^{\tau_\mathrm{V}^\mathrm{Out}}{\zeta(t) x_i(t) \tilde{K}_0(t,\tau_\mathrm{V},s)dt} & \text{if $s > \tau_\mathrm{V}$}\\
    \displaystyle
    \int^s_{\tau_\mathrm{V}^\mathrm{In}}{\zeta(t) x_i(t) \tilde{K}_0(t,\tau_\mathrm{V},s)dt} & \text{if $s < \tau_\mathrm{V}$}
  \end{cases}
.\end{equation}
Finally, we get the derivative of the mean radiation field,
\begin{equation}
  \frac{\delta \bar{\mathcal{J}_{ij}}[x_i,x_j](\tau_\mathrm{V}) }
  {\delta x_i(s)}
   = D_{ij} \zeta(s)
  \left(K(s,\tau_\mathrm{V}) + E_{ij} \times
    \begin{cases}
      \displaystyle
      \int_s^{\tau_\mathrm{V}^\mathrm{Out}}{\zeta(t) x_i(t) \tilde{K}_0(t,\tau_\mathrm{V},s)dt} & \text{if $s > \tau_\mathrm{V}$}\\
      \displaystyle
      \int^s_{\tau_\mathrm{V}^\mathrm{In}}{\zeta(t) x_i(t) \tilde{K}_0(t,\tau_\mathrm{V},s)dt} & \text{if $s < \tau_\mathrm{V}$}
    \end{cases}
  \right)
.\end{equation}

\section{Definition of the variables}
\label{appC}
\begin{center}
\begin{table}[h]
\caption{Physical quantities}   \label{tab:symb1}
\begin{tabular}{ c l l}
\hline\hline
  Symbol & Meaning & units (cgs) \\
  \hline
  $\mu$ & Direction cosine ($\mu=\cos(\theta)$)         & \\
  $T$   & Temperature           & K                     \\
  $n^k$ & Total density of species k            & cm$^{-3}$\\
  $n_i$ & Density in an energy level labeled by $i$ & cm$^{-3}$\\
  $f_i$ & Relative density ($f_i=n_i/n^k$)              & \\
  $x_i$ & Weighted relative density ($x_i= f_i/g_i$)            & \\
  $g_{i}$ & Statistical weight of level $i$ & \\
  $\nu$ & Frequency                     &  Hz   \\
  $I_\nu$ & Specific intensity & erg\,s$^{-1}$\,cm$^{-2}$\,sr$^{-1}$\,Hz$^{-1}$\\
  $\mathcal{J}_\nu$ & Mean intensity    & erg\,s$^{-1}$\,cm$^{-2}$\,Hz$^{-1}$  \\
  $\tau_\nu$ & Optical depth    &  \\
  $\tau_V$ & Optical depth at 500 nm    &  \\
  $S_\nu$ & Source function     & erg\,s$^{-1}$\,cm$^{-2}$\,sr$^{-1}$\,Hz$^{-1}$\\
  $B_\nu$ & Planck function  & erg\,s$^{-1}$\,cm$^{-2}$\,sr$^{-1}$\,Hz$^{-1}$\\
  $\chi_{\nu}$ & Monochromatic extinction & cm$^{-1}$ \\
  $\eta_{\nu}$ & Monochromatic emissivity & erg\,s$^{-1}$\,cm$^{-3}$\,sr$^{-1}$\,Hz$^{-1}$ \\
  $\beta_\nu$ & Escape probability              &  \\
  $A_{ul}$ & Einstein coefficient for spontaneous de-excitation & s$^{-1}$ \\
  $B_{ij}$ & Einstein coefficient for stimulated (de)excitation & erg$^{-1}$\,cm$^2$\,Hz \\
  $C_{ij}$ & Rate for collisional (de)excitation & cm$^3$\,s$^{-1}$\\
  $\phi^{ij}_\nu$ & Normalized line profile of the transition $i\rightarrow j$ & Hz$^{-1}$ \\
  \hline  
  \end{tabular}
\end{table}
\end{center}

\begin{center}
 \begin{table}[!ht]
\caption[]{Functions, operators and distributions}
  \label{tab:symb5}
  \begin{tabular}{ c l l}
\hline\hline
  Functions & Name & Meaning \\
  \hline
    $E_n(x)$ & n$^{\mbox{th}}$ Exponential integral & $ \displaystyle E_n=\int_1^\infty{\frac{e^{-xt}}{t^n}dt}$\\
    $B_\nu(T)$ & Planck function  & $\displaystyle B_\nu(T)=\frac{2h\nu^3}{c^2}\frac{1}{e^{h\nu/k_\text{B}T}-1}$\\
    $\delta(x)$ & Dirac distribution  & $\displaystyle \delta(x)=\begin{cases} +\infty & \mbox{if } x=0 \\ 0 & \mbox{if } x \neq 0. \end{cases}$ \\
    & & $\int_{-\infty}^{+\infty}{\delta(x)dx}=1 $ \\
    $\Pi(x)$ & Gate distribution & $\displaystyle \Pi(x)=H(x+1/2)-H(x-1/2)$\\
    $\Pi_{a,b}(x)$ & Generalized gate distribution (boxcar) & $\displaystyle \Pi_{a,b}(x)=\Pi((x-(b+a)/2(b-a)))$\\
    $\theta(x)$ & Heaviside distribution & $\displaystyle \theta(x)=\begin{cases} 0 & \mbox{if } x<0 \\ 
1 & \mbox{if } x>0 . \end{cases}$ \\
    $\mathscr{G}$ & Green's function  & $\displaystyle \mathcal{L}\mathscr{G}(x,s)=\delta(x-s)$ \\
    $\overline{f}$ & Integration of $f$ weighted by a line profile & $\displaystyle \overline{f}=\int_{-\infty}^{+\infty}{f_\nu\phi_\nu d\nu}$ \\
  \hline  
\end{tabular}
\end{table}%
\end{center}

\end{appendix}

\end{document}